\documentclass[fleqn,12pt]{article}
\usepackage{latexsym}
\usepackage{graphicx}
\usepackage{amsmath}
\usepackage{amssymb}
\usepackage{amsfonts}
\usepackage{epsfig}
\usepackage{textcomp}	
\usepackage{algorithm, algorithmicx, algpseudocode}
\usepackage{mathptmx}
    \usepackage{subcaption}
\usepackage{rotating}

\begin{document}

\title{
A vehicle routing problem for biological sample transportation in healthcare: mathematical formulations and a metaheuristic approach
}
\author{
 Mario Benini \thanks{Dipartimento di Ingegneria dell'Informazione e Scienze Matematiche,
University of Siena, Via Roma, 56, 53100 Siena, Italy, e-mail: mbenini@diism.unisi.it}\and
Paolo Detti \thanks{Dipartimento di Ingegneria dell'Informazione e Scienze Matematiche,
University of Siena, Via Roma, 56, 53100 Siena, Italy, e-mail: detti@dii.unisi.it}  
\and  Garazi Zabalo Manrique de Lara   \thanks{Dipartimento di Ingegneria dell'Informazione e Scienze Matematiche,
University of Siena, Via Roma, 56, 53100 Siena, Italy, e-mail: garazizml@gmail.com}}

\date{}
\maketitle

\begin{abstract}
In this paper, a real-world transportation problem is addressed, concerning the collection and  the transportation of biological sample tubes from sampling points to a main hospital. 
Blood and other biological samples are collected in different centers during morning hours. Then, the samples are transported to the main hospital, for their analysis, by a fleet of vehicles located in geographically distributed depots. 
Each sample has a limited lifetime and must arrive to the main hospital within that time. If a sample cannot arrive to the hospital within the lifetime, either is discarded or must be processed in dedicated facilities called Spoke Centers.

 Two Mixed Integer Linear Programming formulations and an Adaptive Large Neighborhood Search  (ALNS) metaheuristic algorithm have been developed for the problem. Computational experiments on different sets of instances based on real-life data provided by the Local Healthcare Authority of Bologna, Italy, are presented. A comparison  on small instances with the optimal solutions obtained by the formulations  shows the effectiveness of the proposed ALNS algorithm.
 On real-life instances, different batching policies of the samples are evaluated. The results show that the ALNS algorithm is able to find solutions in which all the samples are delivered on time, while in  the real case about the 40\% \cite{ctw20} of the samples is delivered late.


{\bf keywords}:
Vehicle Routing problem, healthcare, lifetime constraints, transfers, Adaptive Large Neighborhood Search, Mixed Integer Linear Programming.
\end{abstract}

\section{Introduction}\label{sec:LUM:intro}
The transportation problem addressed in this paper arises from a real-world healthcare application, concerning the  collection and transportation of biological sample tubes from sampling points to a main laboratory.
More specifically, the problem \cite{Bonadies} arises from the Local Healthcare Authority of Bologna, Italy, and basically consists in collecting and routing blood and other biological samples from different draw centers to the ``Laboratorio Unico Metropolitano'', hereafter called HUB, which is the central laboratory where all samples are analyzed.


Blood and other biological samples are drawn from patients in different centers during morning hours, and are transported to the main hospital to be analysed, by a fleet of vehicles located in geographically distributed depots. Given the small sample dimension, vehicles have no capacity restriction. Each sample has a limited lifetime and must be  delivered to the main hospital within that time. If a sample cannot arrive to the HUB within the lifetime, either is discarded or must be {\em stabilized}. The stabilization process can be performed in dedicated facilities called Spoke Centers or Spokes, and provides the samples with an extra lifetime, useful for their on time delivery. Each sample can be stabilized at most once.
Samples stabilized  at spoke centers can be transferred from a vehicle to another. In fact, vehicle delivering  samples at a spoke center can either wait until the end of the stabilization process or depart after dropping the samples off. Then, the same or another vehicle will  pick up the samples after the end of their stabilization.  

The overall objective of the problem is to minimize the total travelled distance while fulfilling the lifetime requirements of all the samples. In fact, real historical data provided by the Local Healthcare Authority of Bologna, Italy, show that about the 40\% of the samples is delivered late (i.e., after the lifetime) at the main hospital, each day.

The problem has been first addressed in \cite{ods2019,ctw20}. In \cite{ods2019},  the problem is  mathematically modeled, while in \cite{ctw20} a simple metaheuristic  method is presented based on the Adaptive Large Neighborhood Search (ALNS) framework, proposed by Ropke and Pisinger in \cite{ropke2006}.

The main contributions of this paper are listed below.\\ 
$(a)$ A relevant healthcare transportation problem  coming from a real-life application is addressed and solved. More specifically, the problem can be formulated as a new variant of the Vehicle Routing Problem with the following new features:  $i)$ Each transportation request has a limited lifetime and must be delivered within that time to the HUB;  $ii)$ Specific locations exist (i.e., the spokes), devoted to provide extra-lifetimes to the requests;  $iii)$ At the spokes, requests can be transferred from a vehicle to another.
\\ 
$(b)$ Two Mixed Integer Linear Programming (MILP) formulations are developed able to model all the characteristics of the addressed problem.\\
$(c)$ A metaheuristic algorithm based on the Adaptive Large Neighborhood Search framework is proposed, able to solve real-life instances of the problem. The ALNS algorithm is a completely new version of the algorithm proposed in \cite{ctw20}. More specifically, while in \cite{ctw20} spokes were not used during the search process (but only inserted by a post processing procedure at specific algorithm steps), the new algorithm has been designed to handle spokes and sample stabilization at each iteration of the search. Furthermore, the new algorithm has been enriched by a number of new procedures that can be used during the destruction and the repair phases of the solutions. \\
$(d)$ A study is presented to evaluate different grouping policies for managing samples. The aim of this study is to face with the high dimension of the real instances, in which hundreds of samples must be delivered on time each day. The study allows to assess what are the policies yielding the best solutions (i.e., able to deliver all samples on time with smallest traveled distance) in reasonable computational times.  


A computational campaign on different sets of instances based on real-life data is presented. A comparison between the solutions obtained with the MILP formulations and the developed ALNS metaheuristic on small instances, shows the effectiveness of the proposed algorithm, able to find optimal solutions in short time, in almost all the cases.
Furthermore, the algorithm is able to attain solutions in which all samples are delivered on time in real-life instances, while, as stated, real historical data provided by the Local Healthcare Authority of Bologna, show that on average about 40\% of the samples is delivered late.

The paper is organized as follows. In Section \ref{sec:lit}, we review results from the literature.
In Section \ref{sec:LUM:description}, a detailed description of the problem is presented. In Section \ref{sec:milp}, two MILP formulations are presented for the problem. Section \ref{sec:LUM:algorithm} describes the adaptive large neighborhood search algorithm used to solve real-life instances of the problem. The description of the real-life data and the computational results are reported in Section \ref{sec:LUM:res}. Finally, conclusions are gathered in Section \ref{sec:LUM:conclusion}. 

\section{Literature review}\label{sec:lit}

The problem addressed in this paper can be modeled as a variant of the Multi-depot Vehicle Routing Problem (VRP). In the literature, many VRPs and Dial-a-Ride problems have been solved in healthcare contexts for optimizing transportation services \cite{Doerner2008,liu,DPZomega,Zhang}. Furthermore, several works exist that address blood sample collection problems  \cite{Doerner2008,grasas2014improvement, karakoc2017priority, sapountzis1990allocating, csahinyazan2015selective}. In \cite{Doerner2008},  a  transportation problem originating from the blood collection process of the Austrian Red Cross blood program is addressed. The problem is modeled as a vehicle routing problem with multiple interdependent time windows, and solved by a  mixed-integer programming formulation and different heuristics.
Grasas {\em et al.} \cite {grasas2014improvement} consider the problem of sample collection and transportation from different collection points to a core laboratory for testing in Spain.
The problem is modeled as a variant of the capacitated vehicle routing problem with open routes and route length constraints, and a heuristic based on a genetic algorithm is  proposed.
A similar problem is addressed in \cite{csahinyazan2015selective} where a mobile blood collection system is designed and a routing problem is proposed with the aim of transporting blood samples from blood mobile draw centers to the depot. A mathematical model and a 2-stage IP based heuristic algorithm are proposed to solve the problem.
In \cite{karakoc2017priority}, a vehicle routing problem is addressed  for blood transportation between hospitals or donor/client sites. A hybrid meta-heuristic algorithm including genetic algorithms and local search is developed able to reduce the cost and the response time for emergency.  In  \cite{sapountzis1990allocating}, the problem of allocating  units of blood from a regional blood transfusion centre to the hospitals of its area is considered. The problem is formulated as a multi-objective transportation problem.
 
A new feature of the  problem addressed in this paper is the possibility of stabilizing samples in spoke centers to gain extra lifetimes. After the  stabilization, the samples may be taken in charge from a vehicle that is different from the vehicle that delivered them to the spoke. 
In the literature, transportation problems with transfers, i.e., in which a request may be transferred fro one vehicle to another,  have been addressed  in \cite{cortes2010pickup,masson2013adaptive,rais2014}.  In  \cite{rais2014}, the pickup-and-delivery problem in which  transfers are allowed (PDPT) is  addressed, and mixed integer-programming formulations are proposed and evaluated. 
The PDPT  is also addressed in \cite{masson2013adaptive}. The authors propose heuristics capable of efficiently inserting requests through transfer points and embed them into an Adaptive Large Neighborhood Search (ALNS) scheme. The approach is evaluated  on  real-life instances.
Cortes {\em et al.} \cite{cortes2010pickup} address Dial-a-Ride problems where passengers may be transferred from one vehicle to another at specific locations. A mathematical programming formulation is presented and a solution method based on Benders decomposition is proposed.

Unlike in the general scheme of the PDPT, in the problem addressed in this paper only the requests subjected to stabilization at the spoke centers are allowed to be transferred between vehicles.

\section{Problem Description}
\label{sec:LUM:description}
In the problem addressed in this paper, a set of transportation requests, i.e., biological samples, must be carried from draw points to a main hospital (the HUB).
A fleet of vehicles, located in geographically distributed depots is available to perform the transportation requests. The vehicles belong to different organizations and, given the small dimension of the samples, they can be considered with unlimited capacity.
The samples are produced and become available during the opening hours of the draw points. Pick up and delivery operations of the samples require a {\em service time} to load or unload a request. 
Each sample must be delivered to the main hospital within a pre-specified  time span from its withdrawal. In other words, each sample has a lifetime, defining the deadline by which the sample has to be delivered to the HUB.
When a sample cannot be delivered to the hospital within the deadline, it can be {\em stabilized}  to gain an extra lifetime. This stabilization process is performed in geographically distributed spoke centers.
The problem consists in assigning the transportation requests to the vehicles and in finding the routing of each vehicle, in such a way that all samples are delivered on time and the total transportation cost is minimized.

\begin{figure}[htbp]
\begin{center}
\includegraphics[width=0.8\textwidth ]{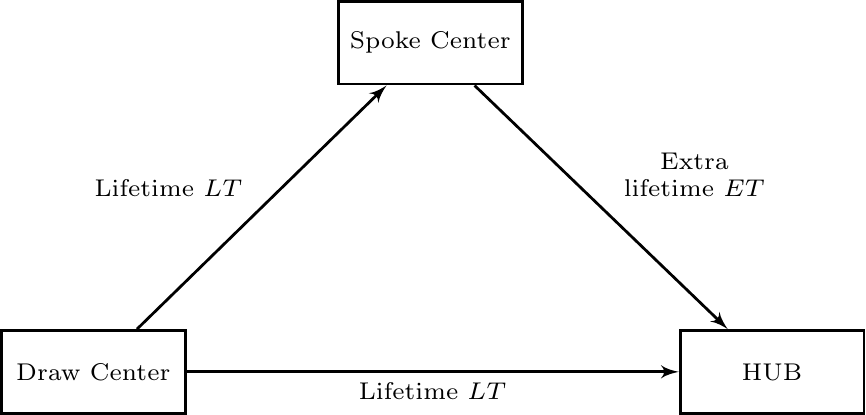}
\caption{Transportation modes of a sample.}
\label{fig:lum:probeldescr}
\end{center}
\end{figure}

Observe that, in a spoke center, a vehicle can depart after dropping a sample off for stabilization, and another vehicle can pass to pick up the sample after the end of the stabilization process. It is important to notice that the transfer of requests between vehicles is allowed only for stabilized requests. In other words, the spoke centers cannot be used for transfer only. Figure \ref{fig:lum:probeldescr} shows a scheme of the two transportation modes of a request (i.e. a sample): either directly from the draw center to the HUB or first from the draw center to a spoke and then to the HUB.

In the remainder of this section, notation is introduced and a  formal definition of the problem is given. Observe that a request, i.e., a sample, must be picked up at the draw center within a time window, defined by the time of its withdrawal from a patient and the closure time of the draw center, and must be delivered to the HUB or to a spoke center within its lifetime.

Let $R$ and $K$ be the sets of transportation requests and vehicles, respectively. Furthermore, let $G=(N,A)$ be a complete directed graph, where $N=\{1,\dots,n,n+1, n+2,\dots,n+s+2, n+s+3, \dots, n+s+m+3\}$ is the node set and $A=\{(i,j): i,j\in N \}$ is the arc set. The nodes in $P=\{1,\dots,n \}$  are the pickup nodes of the transportation request, the node $n+1$, also denoted as $H$, is the delivery node, i.e.,  the main hospital or the HUB. The nodes in $S= \{ n+2,\dots,n+s+2\}$ correspond to the spoke centers and the nodes in $D=\{ n+s+3,\dots,n+s+m+3\}$ are the depots. 
In the following, we denote by $p(r)\in P$ the pickup node of the request $r\in R$ and by $d(k)$ the depot of vehicle $k\in K$.
For each request $r$ in $R$, $LT_r$ is the lifetime of request $r$, and $[e_r, l_r]$ is the related time window. The time window $[e_r, l_r]$ of each request $r\in R$ indicates that $r$ can only be picked up from its pickup location $p(r)$ between time $e_r$ and $l_r$. A vehicle is allowed to arrive at the location $p(r)$ before the start of the time window, but it has to wait $e_r$ to begin the load of the sample. The service time to load/unload requests at each node is $st$.
The lifetime $LT_r$ is the deadline by which request $r$ in $R$ must be delivered either to the HUB or to a spoke center. $ET_r$ is the extra lifetime gained by request $r$ if it is stabilized at a spoke center. Hence, the stabilized request $r$ has to arrive to the HUB (from the spoke) within $ET_r$.  The time needed to perform the stabilization process is the same for all spokes and requests, and is denoted as $stbt$.

 
The problem consists in finding a set of routes on $G$ such that:
\begin{itemize}
\item each route starts and ends at the same depot;
\item each route visits the main hospital right before the arrival to the depot;
\item the service at node $p(r)\in P$ begins in the interval  $[e_r,l_r]$, for each request $r\in R$;
\item each request can be stabilized in a spoke center at most once;
\item all the requests arrive to the main hospital within the lifetimes;
\item the total length of the routes is minimized.
\end{itemize}

\section{Mixed Integer Linear Programming  formulations}\label{sec:milp}

In this section, two time-indexed Mixed Integer Linear Programming  formulations, in the following denoted as $MILP_1$ and $MILP_2$, are presented for the problem under study. Time indexing is used to properly address routes visiting more than once the same spoke center. As an example, multiple visits of a spoke center may occur when a vehicle drops a request at a spoke for stabilization, departs from the spoke and returns to pick up that request after the end of the stabilization. Also requests may visit a spoke multiple times, even if they can be stabilized at most once. For example, a request can pass through a spoke center because the vehicle carrying that request has to drop another sample in that spoke for stabilization.
In order to index the variables with a time component, let's define the time index $t \in TT$, where $TT$ is the time horizon considered.
In the addressed problem is possible to identify two different flows: the flow of vehicles and the flow of requests. The first concerns the route followed by each vehicle from its depot to the HUB and back to its depot. The second is the path of each request from its origin to its final destination. Since the problem allows transshipment of requests at the spoke centers, in general the two flows can be decoupled at spoke centers. So, it is important to keep track of both flows and to make sure that the they are consistent with each other.

In the formulations $MILP_1$ and $MILP_2$, we denote by $loc(i)$ the location of node $i \in N$.

\subsection{Formulation $MILP_1$ } 
Formulation $MILP_1$ employs the following variables: 
\begin{itemize}
\item $x_{ij}^{kt} \in \{0,1\}$ be equal to 1 if vehicle $k$ traverses arc $(i,j)$ arriving in $j$ at $t$ and 0 otherwise;
\item $y_{ij}^{krt} \in \{0,1\}$ is 1 if vehicle $k$ traverses arc $(i,j)$ carrying request $r$ arriving in $j$ at $t$ and 0 otherwise;
\item $z_j^{krt} \in \{0,1\}$ be equal to 1 if vehicle $k$ arrives in spoke $j$ at $t$ to drop request $r$ for stabilization and 0 otherwise;
\item $w_j^{krt} \in \{0,1\}$ be equal to 1 if vehicle $k$ arrives in spoke $j$ at $t$ to pick up request $r$ and 0 otherwise.
\item $A_{r+n}$ arrival time of request $r$ to the HUB.\end{itemize}

The objective function is to minimize the total traveled distance:
\begin{equation}
    \min \sum_{t \in TT}\sum_{k \in K}\sum_{(i,j) \in A} d_{ij}x_{ij}^{kt}
\end{equation}

where $d_{ij}$ is the travel distance between nodes $i$ and $j$ of $N$.

The constraints of the model can be divided into three classes:
$(i)$ {\em Flow Constraints}, describing both the flow of requests and the flow of vehicles;
$(ii)$  {\em Spoke Constraints}, describing how the spoke nodes work;
$(iii)$  {\em Time Constraints}, imposing the timing restrictions of the problem and the compliance of visit time of the nodes.\\

The Flow Constraints read as follow.
\begin{eqnarray}
    \sum_{t \in TT} \sum_{ j \in N, j \neq d(k) } x^{kt}_{d(k)j} \leq 1, &&\forall k \in K
 \label{eq:1}\\
    \sum_{t \in TT} x^{kt}_{Hd(k)} = \sum_{t \in TT} \sum_{i \in N, i \neq H} x^{kt}_{iH}, &&\forall k \in K
 \label{eq:2}\\
     \sum_{t \in TT} x^{kt}_{ij} +  \sum_{t \in TT} x^{kt}_{ji} \leq 1, \forall i \in D, \forall j \in S, \forall k \in K&&
 \label{eq:3}\\
    \sum_{t \in TT} \sum_{ j \in N, j \neq d(k) } x^{kt}_{d(k)j} = \sum_{t \in TT} \sum_{ j \in N, j \neq d(k) } x^{kt}_{jd(k)}, && \forall k \in K
 \label{eq:4}\\
    \sum_{t \in TT} \sum_{ j \in N, j \neq i} x^{kt}_{ij} - \sum_{t \in TT} \sum_{ j \in N, j \neq i} x^{kt}_{ji} = 0,&& \forall k \in K, \forall i \in N
 \label{eq:5}\\
    x^{kt}_{p(j)p(i)} = 0,  \forall k \in K, \forall t \in TT, \forall j,i \in R : loc(p(i)) = loc(p(j)), e_i < e_j&&
 \label{eq:6}\\ 
    \sum_{t \in TT} \sum_{k \in K} \sum_{j \in N, j \neq i} x^{kt}_{ij} = 1, &&\forall i \in P
 \label{eq:7}\\
    \sum_{t \in TT} \sum_{k \in K} \sum_{j \in N, j \neq p(r)} y^{krt}_{p(r)j} = 1, &&\forall r \in R
 \label{eq:8}\\ 
    \sum_{t \in TT} \sum_{k \in K} \sum_{i \in N, i \neq H} y^{krt}_{iH} = 1, &&\forall r \in R \label{eq:9}\\
    \sum_{t \in TT} \sum_{k \in K} \sum_{j \in N, j \neq i} y^{krt}_{ij} - \sum_{t \in TT} \sum_{k \in K} \sum_{j \in N, j \neq i} y^{krt}_{ji} = 0, &&\forall r \in R, \forall i \in S \label{eq:10}\\
    \sum_{t \in TT} \sum_{j \in N, j \neq i} y^{krt}_{ij} - \sum_{t \in TT} \sum_{j \in N, j \neq i} y^{krt}_{ji} = 0, \forall r \in R, \forall k \in K, \forall i \in P \setminus p(r)&& \label{eq:11}\\
    \sum_{t \in TT} \sum_{k \in K} \sum_{j \in N, j \neq p(r)} y^{krt}_{jp(r)} &=& 0, \forall r \in R \label{eq:12}\\
    \sum_{t \in TT} \sum_{k \in K} \sum_{j \in N, j \neq p(r)} y^{krt}_{p(r)j} - \sum_{t \in TT} \sum_{k \in K} \sum_{j \in N, j \neq p(r)} y^{krt}_{jp(r)} = 1, &&\forall r \in R \label{eq:13}\\
    y^{krt}_{ij} \leq x^{kt}_{ij}, \forall r \in R, \forall k \in K, \forall (i,j) \in A, \forall t \in TT &&\label{eq:14}
\end{eqnarray}
Constraints (\ref{eq:1}) ensure that each vehicle leaves at most once from its depot, while Constraints (\ref{eq:2}) state that a vehicle must come back to its depot after visiting the HUB. Constraints (\ref{eq:3}) forbid a vehicle to go directly from its depot to a spoke and come back to its depot immediately after. Constraints (\ref{eq:4}) ensure that each vehicle must return in its depot. Vehicle flow conservation is imposed by equalities (\ref{eq:5}), while Constraints (\ref{eq:6}) are used to forbid sub-cycles between pickup nodes. Constraints (\ref{eq:7}) ensure, together with (\ref{eq:6}), that each pickup node must be visited by exactly one vehicle. Constraints (\ref{eq:8}) and (\ref{eq:9}) ensure that each request is served exactly once. Constraints (\ref{eq:10}) maintain the request flow conservation at the spoke nodes allowing requests to switch from one vehicle to another. Constraints (\ref{eq:11})-(\ref{eq:13}) ensure the request flow conservation at the pickup nodes requiring that every vehicle visiting a pickup node with some requests must leave carrying those requests and the one related to the visited node.
Finally, Constraints (\ref{eq:14}) link variables $x$ and $y$.

Now, the Spoke Constraints are presented.
\begin{eqnarray}
    \sum_{t \in [e_r,e_r+LT_r]} \sum_{k \in K} \sum_{ j \in S} z^{krt}_{j} \leq 1, \forall r \in R
\label{eq:15}\\
    z^{krt}_{j} \leq \sum_{ i \in N, i \neq j} y^{krt}_{ij}, \forall j \in S, \forall k \in K, \forall r \in R, \forall t \in [e_r,e_r+LT_r]
\label{eq:16}\\
    \sum_{t \in [e_r,l_r+LT_r]} \sum_{k \in K}  z^{krt}_{j} = \sum_{t \in [e_r,e_r+LT_r+stbt+ET_r]} \sum_{k \in K} w^{krt}_{j}, \forall j \in S, \forall r \in R
\label{eq:17}\\
x^{kt}_{ij} \leq \sum_{ r \in R} z^{krt}_{j} + \sum_{ \tau=t+st}^{TT} \sum_{ r \in R} w^{kr\tau}_{j}, \forall j \in S, \forall k \in K, \forall i \in N, i \neq j, \forall t \in TT
\label{eq:18}\\
\sum_{ i \in N, i \neq j} y^{kr(t+t_{ji})}_{ji} \ge w^{krt}_{j} , \forall r \in R, \forall j \in S, \forall k \in K, \forall t\in [e_r,e_r+LT_r+stbt+ET_r]
\label{eq:19}\\
z^{krt}_{j} \leq \sum_{ q \in K} \sum_{ \tau=t+st+stbt+st}^{e_r+LT_r+stbt+ET_r} w^{qr\tau}_{j}, \forall j \in S, \forall k \in K, \forall r \in R, \forall t \in [e_r,e_r+LT_r]
\label{eq:20}\\
w^{qrt}_{j} \leq \sum_{ q \in K} \sum_{ \tau=e_r}^{t-st-stbt} z^{qr\tau}_{j}, \forall j \in S, \forall k \in K, \forall r \in R, \forall t \in [e_r,e_r+LT_r+stbt+ET_r]
\label{eq:21}\\
\sum_{i \in N, i \neq j}\sum_{\tau \geq t+st+t_{ji}} x_{ji}^{k\tau} \geq z_j^{krt} - M[(1-z_j^{krt})] \nonumber\\ \forall r \in R, \forall k \in K, \forall j \in S, \forall t\in [e_r,e_r + LT_r]
\label{eq:22}\\
\sum_{i \in N, i \neq j}\sum_{\tau=t+st+t_{ji}}^{t_2-1} x_{ji}^{k\tau} \geq 1 - M[(1-\sum_{l \in N,l\neq j}x_{lj}^{kt}) + (1- \sum_{l \in N,l\neq j}x_{lj}^{kt_2}],  \forall k \in K, \nonumber\\ \forall j \in S, \forall t,t_2 \in TT, t>e_r, t_2>t
\label{eq:23}\\
\sum_{i \in N, i\neq j}\sum_{\tau\geq t+st+t_{ji}} y_{ji}^{kr\tau} \geq \sum_{i \in N, i\neq j} y_{ij}^{krt} - z_j^{krt}, \forall j \in S, \forall r \in R, \forall k \in K, \forall t \in TT, t>e_r
\label{eq:24}\\
\sum_{t\in TT}\sum_{j\in S} z_{j}^{krt}\le \sum_{h\in N}\sum_{t\in TT} x_{hi(r)}^{kt}, \forall r \in R, i \in P
\label{eq:25}
\end{eqnarray}

Constraints (\ref{eq:15}) impose that each request can be stabilized in a spoke at most once. Constraints (\ref{eq:16}) link variables $y$ and $z$, while Constraints (\ref{eq:17}) state that, if a request is left in a spoke node for the stabilization, then it must be picked up from that spoke node. Constraints (\ref{eq:18}) ensure that a vehicle can visit a spoke node only if it has either to leave a request for stabilization or to withdraw a stabilized request. Constraint (\ref{eq:19}) links the variables $y$ and $w$ while Constraints (\ref{eq:20})-(\ref{eq:21}) link the timing between the $z$ and $w$ variables. Constraints (\ref{eq:22})-(\ref{eq:23}) contribute to define the exit time of a vehicle from a spoke node and to manage the multiple visits at a spoke node by imposing that each vehicle has to depart from a spoke node within the time interval between two different visits of the node. Constraints (\ref{eq:24}) impose that a request transfer at a spoke center is not allowed unless the request is left for stabilization. Lastly, Constraints (\ref{eq:25}) state that a request can be left at a spoke node for stabilization only by the vehicle that loaded it from its pickup node.

Finally, the Timing Constraints are reported in the following.
\begin{eqnarray}
 e_r\le B_p(r) \le l_r,\forall r \in R  \quad
\label{eq:26}\\
B_j\ge\sum_{ k \in K}\sum_{ t=t_{d(k)}}^{TT}\sum_{ i \in N, i \neq j} t x^{kt}_{ij},  \forall j \in P
\label{eq:27}\\
tx_{ij}^{kt} \geq B_i + st + t_{ij} - M(1-x_{ij}^{kt}), \forall t \in TT, t>min_{r}{e_r}, \forall k \in K, \forall i \in P, \forall j\in N, loc(i) \neq loc(j)& &
\label{eq:28}\\
tx_{ij}^{kt} \leq B_i + st + t_{ij} + M(1-x_{ij}^{kt}), \forall t \in TT, t>min_{r}{e_r},\forall i \in P, \forall j\in N, loc(i) \neq loc(j)&&
\label{eq:29}\\
tx_{ij}^{kt} \geq B_i - M(1-x_{ij}^{kt}), \forall t \in TT, t>min_{r}{e_r},\forall k \in K, \forall i \in P, \forall j\in N, loc(i) = loc(j)
\label{eq:30}\\
tx_{ij}^{kt} \leq B_i + M(1-x_{ij}^{kt}), \forall t \in TT, t>min_{r}{e_r},\forall i \in P, \forall j\in N, loc(i) = loc(j)
\label{eq:31}\\
A_{r+n}= \sum_{t\in TT, t>e_r} \sum_{k \in K} \sum_{i \in N, i \neq H} t y^{krt}_{iH} \forall r \in R
\label{eq:32}\\
A_{r+n} +st - e_r \leq LT_r(1-\sum_{t=e_r}^{e_r+LT_r} \sum_{k \in K} \sum_{j \in S} z_j^{krt}) +M\sum_{t=e_r}^{e_r+LT_r} \sum_{k \in K} \sum_{j \in S} z_j^{krt}, \forall r \in R
\label{eq:33}\\
A_{r+n} +st - \sum_{t=e_r}^{e_r+LT_r} \sum_{k \in K} \sum_{j \in S} (t+st+stbt) z_j^{krt} \leq \nonumber\\ M(1-\sum_{t=e_r}^{e_r+LT_r} \sum_{k \in K} \sum_{j \in S} z_j^{krt}) +ET_r\sum_{t=e_r}^{e_r+LT_r} \sum_{k \in K} \sum_{j \in S} z_j^{krt}, \forall r \in R
\label{eq:34}\\
\sum_{t=e_r}^{e_r+LT_r} \sum_{k \in K} \sum_{j \in S} (t+st) z_j^{krt} - e_r \leq LT_r, \forall r \in R
\label{eq:35}
\end{eqnarray}
\\
\\
Time windows compliance is ensured  by Constraints (\ref{eq:26}), while Constraints (\ref{eq:27}) state that the loading service at a pickup node can  start after the arrival of a vehicle. Constraints (\ref{eq:28})-(\ref{eq:31}) ensure the correct timing between two pickup nodes while Constraints (\ref{eq:32}) define the arrival at the HUB variables. Compliance of lifetimes is ensured by Constraints (\ref{eq:33}) when a request is not stabilized at a spoke, while Constraints (\ref{eq:34})-(\ref{eq:35}) ensure the lifetime and extra lifetime compliance when a request is stabilized at a spoke node.

\subsection{Formulation $MILP_2$}
A different formulation with a smaller number of variables, denoted as $MILP_2$, can be obtained by dropping variables $y$ from formulation $MILP_1$. 
Variables $x$ and $y$ are used in $MILP_1$ to represent the two different flows of requests and vehicles  and to make them match correctly. However, with the only use of variables $x$ is still possible to track the flow of requests at least until they arrive at a spoke node. In fact, if $x_{ij}^{kt}$ is equal to 1 for a certain $i\in P$ (or $j \in P$) it means that the request related to the node $i$ (or $j$) will be on board of vehicle $k$ at least until the next spoke node is visited by that vehicle. If the vehicle does not visit a spoke node after the visit of node $i$ ($j$) then the related request flow will certainly be the same as the vehicle $k$ flow until the HUB. Variables $z$ also allow to determine the spoke node in which a vehicle leaves a request for stabilization. Hence, it is possible to track the flow of a certain request $r$ by the $x$ variables, until the vehicle carrying $r$ reaches the spoke node in which the request is stabilized. Furthermore, the use of the variables $w$ allows us to track the requests flow also after the spoke node, without the need of the $y$ variables. In fact, if $w_j^{krt}$ is equal to 1, then vehicle $k$ picks up request $r$ at the spoke node $j$ after the stabilization and, since a request can switch vehicle only when stabilized, it follows that from time $t$ on the flow of request $r$ will match entirely to the flow of vehicle $k$. 

In the following the $MILP_2$ model is presented.
As for $MILP_1$, also in $MILP_2$ the constraints of the model can be divided into three classes.
The flow constraints of the model are reported in the following.

\begin{eqnarray}
    \sum_{t \in TT} \sum_{ j \in N, j \neq d(k) } x^{kt}_{d(k)j} \leq 1, \forall k \in K
 \label{MILP3_1}\\
    \sum_{t \in TT} x^{kt}_{Hd(k)} = \sum_{t \in TT} \sum_{i \in N, i \neq H} x^{kt}_{iH}, \forall k \in K
 \label{MILP3_2}\\
     \sum_{t \in TT} x^{kt}_{ij} +  \sum_{t \in TT} x^{kt}_{ji} \leq 1, \forall i \in D, \forall j \in S, \forall k \in K
 \label{MILP3_3}\\
    \sum_{t \in TT} \sum_{ j \in N, j \neq d(k) } x^{kt}_{d(k)j} = \sum_{t \in TT} \sum_{ j \in N, j \neq d(k) } x^{kt}_{jd(k)}, \forall k \in K
 \label{MILP3_4}\\
    \sum_{t \in TT} \sum_{ j \in N, j \neq i} x^{kt}_{ij} - \sum_{t \in TT} \sum_{ j \in N, j \neq i} x^{kt}_{ji} = 0, \forall k \in K, \forall i \in N
 \label{MILP3_5}\\
    x^{kt}_{p(j)p(i)} = 0, \forall k \in K, \forall t \in TT, \forall j,i \in R : loc(p(i)) = loc(p(j)), e_i < e_j
 \label{MILP3_6}\\
    \sum_{t \in TT} \sum_{k \in K} \sum_{j \in N, j \neq i} x^{kt}_{ij} = 1, \forall i \in P
 \label{MILP3_7}\\
    \sum_{t \in TT} \sum_{j \in N, j \neq i} x^{kt}_{jH} \leq 1, \forall k \in K
 \label{MILP3_8}
\end{eqnarray}
Constraints (\ref{MILP3_1})-(\ref{MILP3_7}) are the same as Constraints (\ref{eq:1})-(\ref{eq:7}) of $MILP_1$ while (\ref{MILP3_8}) states that a vehicle can go to the HUB at most once. 
\\
\\
Now, the spoke constraints of the $MILP_2$ model are presented.
\\
\begin{eqnarray}
    \sum_{t \in [e_r,e_r+LT_r]} \sum_{k \in K} \sum_{ j \in S} z^{krt}_{j} \leq 1, \forall r \in R
\label{MILP3_9}\\
    z^{krt}_{j} \leq \sum_{ i \in N, i \neq j} x^{kt}_{ij}, \forall j \in S, \forall r \in R, \forall k \in K, \forall t \in [min_{r\in R}e_r,max_{r\in R}e_r+max_{r\in R}LT_r]
\label{MILP3_10}\\
    \sum_{t \in [e_r,e_r+LT_r]} \sum_{k \in K}  z^{krt}_{j} = \sum_{t \in [e_r,e_r+LT_r+stbt+ET_r]} \sum_{k \in K} w^{krt}_{j}, \forall j \in S, \forall r \in R
\label{MILP3_11}\\
x^{kt}_{ij} \leq \sum_{ r \in R} z^{krt}_{j} + \sum_{ \tau=t+st}^{TT} \sum_{ r \in R} w^{kr\tau}_{j}, \forall j \in S, \forall k \in K, \forall i \in N, i \neq j, \forall t \in TT
\label{MILP3_12}\\
\sum_{ i \in N, i \neq j} x^{kr(t+t_{ji})}_{ji} \ge w^{krt}_{j} , \forall j \in S, \forall k \in K, \forall r \in R \nonumber \\ \forall t\in [min_{r\in R}e_r,max_{r\in R}e_r+max_{r\in R}LT_r+st+stbt+max_{r\in R}ET_r]
\label{MILP3_13}\\
z^{krt}_{j} \leq \sum_{ q \in K} \sum_{ \tau=t+st+stbt+st}^{e_r+LT_r+stbt+ET_r} w^{qr\tau}_{j}, \forall j \in S, \forall k \in K, \forall r \in R, \forall t \in [e_r,e_r+LT_r]
\label{MILP3_14}\\
w^{qrt}_{j} \leq \sum_{ q \in K} \sum_{ \tau=e_r}^{t-st-stbt} z^{qr\tau}_{j}, \forall j \in S, \forall k \in K, \forall r \in R, \forall t \in [e_r,e_r+LT_r+stbt+ET_r]
\label{MILP3_15}\\
\sum_{i \in N, i \neq j}\sum_{\tau \geq t+st+t_{ji}} x_{ji}^{k\tau} \geq z_j^{krt} - M[(1-z_j^{krt})] \nonumber\\ \forall r \in R, \forall k \in K, \forall j \in S, \forall t\in [e_r,e_r + LT_r]
\label{MILP3_16}\\
\sum_{i \in N, i \neq j}\sum_{\tau=t+st+t_{ji}}^{t_2-1} x_{ji}^{k\tau} \geq 1 - M[(1-\sum_{l \in N,l\neq j}x_{lj}^{kt}) + (1- \sum_{l \in N,l\neq j}x_{lj}^{kt_2}],  \forall k \in K, \nonumber\\ \forall j \in S, \forall t,t_2 \in TT, t>e_r, t_2>t
\label{MILP3_17}\\
\sum_{t\in TT}\sum_{j\in S} z_{j}^{krt}\le \sum_{h\in N}\sum_{t\in TT} x_{hi(r)}^{kt}, \forall r \in R, i \in P
\label{MILP3_18}\\
\sum_{\tau \leq t+st+min_j{t_{p(r)j}}} z_j^{kr\tau} \leq M(1- \sum_{i \in N, i \neq p(r)}x_{ip(r)}^{kt}), \forall k \in K, \forall r \in R, \nonumber\\ 
\forall t \in[e_r,e_r+LT_r+stbt+ET_r-st]
\label{MILP3_19}\\
\sum_{i \in N,i\neq H}\sum_{t \in TT}x_{iH}^{kt} \geq \sum_{i \in N,i\neq H}\sum_{t \in TT}x_{ip(r)}^{kt} - \sum_{j \in S,i\neq H}\sum_{t \in TT}z_j^{krt} + \sum_{j \in S,i\neq H}\sum_{t \in TT}w_j^{krt}, \forall r\in R, \forall k\in K
\label{MILP3_20}
\end{eqnarray}
Constraints (\ref{MILP3_9}), (\ref{MILP3_11}) and (\ref{MILP3_12}) are the same as (\ref{eq:15}), (\ref{eq:17}), (\ref{eq:18}) of $MILP_1$, while Constraints (\ref{MILP3_10}) and (\ref{MILP3_13}) replace (\ref{eq:16}) and (\ref{eq:19}), respectively, ensuring consistency between the flow of vehicles and the download/withdraw of requests at a spoke node. Constraints (\ref{MILP3_14})-(\ref{MILP3_18}) are the same as (\ref{eq:20})-(\ref{eq:23}) and (\ref{eq:25}) of $MILP_1$. Constraints (\ref{MILP3_19}) impose the minimum time needed for a request to be left at a spoke node after its withdrawal from its pickup location. Lastly, Constraints (\ref{MILP3_20}) force a vehicle to reach the HUB if it has withdrawn a request at its pickup location and the request is not stabilized, or if it picks up a request after stabilization at a spoke node.

Finally, the Timing Constraints of the $MILP_2$ model are:
\begin{eqnarray}
    e_r\le B_p(r) \le l_r,  \forall r \in R
\label{MILP3_21}\\
B_j\ge\sum_{ k \in K}\sum_{ t=t_{d(k)}}^{TT}\sum_{ i \in N, i \neq j} t x^{kt}_{ij},  \forall j \in P
\label{MILP3_22}\\
tx_{ij}^{kt} \geq B_i + st + t_{ij} - M(1-x_{ij}^{kt}), \forall t \in TT, t>min_{r}{e_r}, \forall k \in K, \forall i \in P, \forall j\in N, loc(i) \neq loc(j)
\label{MILP3_23}\\
tx_{ij}^{kt} \leq B_i + st + t_{ij} + M(1-x_{ij}^{kt}), \forall t \in TT, t>min_{r}{e_r},\forall i \in P, \forall j\in N, loc(i) \neq loc(j)
\label{MILP3_24}\\
tx_{ij}^{kt} \geq B_i - M(1-x_{ij}^{kt}), \forall t \in TT, t>min_{r}{e_r},\forall k \in K, \forall i \in P, \forall j\in N, loc(i) = loc(j)
\label{MILP3_25}\\
tx_{ij}^{kt} \leq B_i + M(1-x_{ij}^{kt}), \forall t \in TT, t>min_{r}{e_r},\forall i \in P, \forall j\in N, loc(i) = loc(j)
\label{MILP3_26}\\
A_{r+n} \geq \sum_{i\in N}\sum_{t\in TT} tx_{iH}^{kt} - M(1-\sum_{h\in N}\sum_{t\in TT} x_{hi(r)}^{kt}) -M \sum_{q\in K}\sum_{t\in TT}\sum_{j\in S} z_{j}^{qrt}, \forall r \in R, \forall k \in K
\label{MILP3_27}\\
A_{r+n} \geq \sum_{i\in N}\sum_{t\in TT} tx_{iH}^{kt} -M (1-\sum_{t\in TT}\sum_{j\in S} w_{j}^{krt}), \forall r \in R, \forall k \in K
\label{MILP3_28}\\
A_{r+n} +st - e_r \leq LT_r(1-\sum_{t=e_r}^{e_r+LT_r} \sum_{k \in K} \sum_{j \in S} z_j^{krt}) +M\sum_{t=e_r}^{e_r+LT_r} \sum_{k \in K} \sum_{j \in S}z_j^{krt}, \forall r \in R
\label{MILP3_29}\\
A_{r+n} +st - \sum_{t=e_r}^{e_r+LT_r} \sum_{k \in K} \sum_{j \in S} (t+st+stbt) z_j^{krt}) \leq \nonumber\\ M(1-\sum_{t=e_r}^{e_r+LT_r} \sum_{k \in K}
\sum_{j \in S} z_j^{krt}) +ET_r\sum_{t=e_r}^{l_r+LT_r} \sum_{k \in K} \sum_{j \in S} z_j^{krt}, \forall r \in R
\label{MILP3_30}\\
\sum_{t=e_r}^{e_r+LT_r} \sum_{k \in K} \sum_{j \in S} (t+st) z_j^{krt} - e_r \leq LT_r, \forall r \in R
\label{MILP3_31}
\end{eqnarray}
Constraints (\ref{MILP3_21})-(\ref{MILP3_26}) are the same as (\ref{eq:26})-(\ref{eq:31}) of $MILP_1$. Constraints (\ref{MILP3_27})-(\ref{MILP3_28}) replace (\ref{eq:32}) and define the arrival at the HUB of a request on the basis of the vehicle that is carrying the request. Lastly, (\ref{MILP3_29})-(\ref{MILP3_31}) are the same as (\ref{eq:33})-(\ref{eq:35})

\section{An Adaptive Large Neighborhood Search algorithm for the problem}
\label{sec:LUM:algorithm}

In this section, the ALNS algorithm developed for the addressed problem is presented.
The  framework of our algorithm is mainly based on the one proposed in \cite{ropke2006}, in which  specific developments for the problem under study have been introduced. However, while  in \cite{ropke2006} the algorithm only accepts feasible solutions, in our approach we allow the exploration of unfeasible solutions, too. Infeasibilities are properly included into the objective function as penalty terms. In practice, 
in real-cases instances the existence of a feasible solution can not be guaranteed a priori.
Furthermore, the strict lifetime requirements,
the existence of different transportation modes of the requests, the possibility of transfers, as well as the possibility of multiple visits at the spoke centers make hard even the detection of a feasible solution.  

 The developed ALNS algorithm has two main steps. First an initial solution, $s_0$, is generated not necessarily feasible, through different heuristics. Afterwards the solution is iteratively destroyed and repaired until a maximum number of iterations is reached. At each iteration a solution (not necessarily feasible) is accepted under given acceptance criteria. If the criteria are satisfied, the solution is kept and the algorithm continues with the next iteration. The process is repeated until the maximum number of iterations is reached. 
 
In Algorithm \ref{algo:alns},  a general scheme of the ALNS is reported. In what follows, the main building blocks of the algorithm are described into detail. 
\begin{algorithm}
\caption{ALNS Algorithm}\label{algo:alns}
{\scriptsize
\begin{tabbing}
Generate the initial solution $s_{0}$;\\
Set $s_{best}=s_0$\; and \;$s=s_0$ \; $f(s_{bestFeas})=+\infty$\\
{\bf {Repeat}}\\
\quad $s'\leftarrow Destroy(s)$\\
\quad  $s'\leftarrow Repair(s')$\\
\quad {\bf{If }}  $s'$ is feasible and $f(s')<f(s_{bestFeas})$\\
  \quad\quad Set $s_{bestFeas}:=s' $\\
 \quad {\bf{If }}  $f(s')<f(s_{best})$\\
  \quad\quad Set $s_{best}:=s' $\\
  \quad {\bf{If }}  $Accept(s')=true$\\
  \quad\quad $s=s' $\\


{\bf{Until}}  $It_{max}$ iterations are reached \\
return $s_{best}$ and $s_{bestFeas}$.
 
\end{tabbing}
}
\end{algorithm}

\subsection{Evaluation Function and  route length minimization} \label{sec:lum:evfunc}
As already stated in the above section, during the algorithm  we allow the exploration of infeasible solutions in order to facilitate the search in the solution space. More precisely, the violation of the two main sets of constraints of the problem is allowed, i.e., the constraints on the lifetimes (i.e., deadlines)  and on the time windows of the requests. The violations of these constraints are weighted in the evaluation function by penalty positive coefficients.

Given a solution $s$ of the problem, the evaluation function tackled by ALNS has two main components and is equal to $f(s)=f_1(s)+f_2(s)$. The function $f_1(s)$ corresponds to the total traveled  distance by all the routes. The term $f_2(s)$ is a penalization component of the form $f_2(s)=\alpha t(s)+\beta w(s)$, where $t(s)$ and $w(s)$ represent the total violation of solution $s$ with respect to lifetimes and time windows constraints, respectively, and $\alpha$ and $\beta$ are positive penalty coefficients. In more detail, violations are calculated as in \eqref{eq:viol} and \eqref{eq:viol1}. In solution $s$, given a  request $i \in P$ and its related pick up time window $[e_i,l_i]$, we denote by 
$p_i$ and $d_i$ the times in which the request is picked up from its pickup location and is delivered to the main hospital, respectively. Let  $\delta(i)$ be 1 if $i$ is stabilized in a spoke Center in $s$ and 0 otherwise. If $i$ is stabilized, then let $sa_i$  be the time in which $i$ arrives at the spoke center and let $se_i$ be the time in which $i$ ends the stabilization process.

Then 
\begin{eqnarray}
t(s) &=& \displaystyle \sum_{i\in P} ((d_i-e_i-LT_i)^+ (1-\delta(i))+  (sa_i-e_i-LT_i)^+ \delta(i)+ (d_i-se_i-ET_i)^+ \delta(i))\label{eq:viol}\\
w(s) &=&\displaystyle \sum_{i\in P}(p_i-l_i)^+.\label{eq:viol1}
\end{eqnarray}

(Where $(a)^+=a$ if $a>0$ and 0 otherwise.)
\medskip 

In $f(s)$,  coefficients $\alpha$ and $\beta$ are initially set to given values $\alpha_0$ and $\beta_0$ respectively. Then, at each iteration, $\alpha$ and $\beta$ are increased (decreased) if  lifetime or time windows constrains are  violated (not violated) in solution $s$. 


%
%
%
%

\subsection{Destruction Heuristics}\label{sec:lum:destr}
In the ALNS algorithm, destruction operations are performed by six different heuristics. These heuristics remove a given number  of requests from a given solution $s$. All the destruction heuristics have been designed to tackle with spoke centers. When a request that is stabilized in a spoke center is  removed from the current solution $s$, all the information related to the pair request/spoke is deleted from $s$ and all involved routes are updated.
Three of the destruction heuristics have already been introduced in the literature and have been adapted for the problem under study. They are the {\em Worst Removal}, the {\em Random Removal} \cite{ropke2006} and the {\em Related Removal} \cite{Shaw}, and are briefly described in the following.
\begin{itemize}
\item The Worst Removal heuristic iteratively removes requests that cause the biggest detour in the current solution. In particular, the requests are removed in a semi-random way. The probability for a request to be removed from the solution is proportional to the decrease in the objective function produced by the removal of that request. The randomization is regulated by a parameter $p\geq 1$, the smaller $p$ is, the more randomness is allowed.
\item The Random Removal heuristic randomly selects the requests to remove.
\item The Related Removal (or Shaw Removal) heuristic aims to remove requests that are more related with each other in order to give them the chance to be re-routed more efficiently. A specific measure of relatedness between requests have been developed ad-hoc for the problem under study and is  described into detail in Section \ref{sec:relrem}.
\end{itemize}

The other three heuristics were developed ad-hoc for the problem and are denoted as {\em Vehicle Removal},  {\em Spoke Removal} and  {\em Late Requests Removal}. They are briefly described in the following, and in more detail in Sections \ref{sec:veirem}--\ref{sec:laterem}.
\begin{itemize}
\item The Vehicle Removal heuristic removes all the requests of a randomly selected vehicle.
\item The Spoke Removal heuristic removes all the requests stabilized at a randomly selected spoke center.
\item The Late Requests Removal removes the requests with the higher lifetime violations.
\end{itemize}

\subsubsection{Related Removal and relatedness function}\label{sec:relrem}
As suggested by Shaw \cite{Shaw}, the removal of requests that are very different from each other may not help in improving the solution when re-inserting them. In fact the solution obtained after the re-insertion will probably be close to the previous solution, because the requests will likely be re-inserted in their original positions if not even in worse positions. Therefore, the general idea of this heuristic is to remove requests that are somehow related to each other, in order to give them the chance to be re-routed efficiently. 
Let $R(i,j)$ be the \textit{relatedness function} of requests $i$ and $j$. The bigger $R$ is the more related are the two requests. For the problem under study, we developed an ad-hoc measure of relatedness between a pair of requests, which takes into account three pieces of information: the distance between their pick-up locations, possible overlaps of their time windows, and if they are routed on the same vehicle in the current solution or not. More precisely, the relatedness function employed in the Related Removal heuristic is computed as $R(i,j)=1/(D_{ij} + E_{ij} + V_{ij})$, where $D_{ij}$ is the distance between request $i$ and request $j$ (normalized according to the maximum distance between locations), $E_{ij}=1/(100(l_i-e_j))$ if $l_i-e_j \geq 0$ and $E_{ij}=1+1/(100(l_i-e_j))$ if $l_i-e_j \le 0$, and $V_{ij}$ is equal to 1 if the requests are on the same vehicle and 0 otherwise.
The value $l_i-e_j$ is a measure of the overlap between the time windows of requests $i$ and $j$. If the two time windows overlap with each other, then $l_i-e_j \geq 0$ and $E_{ij}$ is smaller, otherwise $l_i-e_j \le 0$ and $E_{ij}$ is bigger. Recall that, $l_i$ is the end of the time window of request $i$, and  $e_j$ is the beginning of the time window of request $j$. 
Hence, $R(i,j)$ is bigger when $D_{ij}$, $E_{ij}$ and $V_{ij}$ are smaller, implying that the requests are more related. In fact, we consider two requests $i$ and $j$ more related when they have small values of $D_{ij}$, $E_{ij}$ and $V_{ij}$, meaning that their pick-up locations are close together, there is overlap between their time windows and they are not on the same vehicle. So, requests with higher relatedness are those currently in different vehicles, but that could be more appropriately assigned to the same route.

\subsubsection{Vehicle Removal}\label{sec:veirem}
In the problem under study is often possible to reduce the number of used vehicles by routing some requests through a spoke center. In fact, since the stabilization increases the lifetimes of requests, a single vehicle is able to transport a bigger number of requests.
The Vehicle Removal heuristic simply  removes a route from the solution. In this way, the requests carried on the removed vehicle have the chance to be re-routed with another vehicle, possibly through a spoke center. In particular, the heuristic deletes a route from the solution by removing all the requests in that route. The route with the largest total violation of the time windows and lifetime constraints is removed. If all routes in the current solution are feasible, then a random route is removed. Observe that, even if the requests will not be reassigned in already existing routes, by removing the route with the largest violations, after the re-insertion of the removed requests the solution will likely be better than before. 
Finally, note that the number of requests removed by the heuristic is not known \textit{a priori} but depends on the route selected.

\subsubsection{Spoke Removal}\label{sec:sprem}
This heuristic is inspired from a neighborhood for a pickup and delivery problem with transfer proposed in \cite{masson2013adaptive}. The idea is to remove all the requests using a given spoke center, in order to give them the chance to be re-routed through a different spoke center or without stabilization. We first select a random spoke among the active ones in the solution. If the number of requests using that spoke center is less than or equal to the number of requests to be destroyed, all these requests are removed from the solution. In this way the spoke center is also deleted from the current solution. If the number of requests using the selected spoke is greater than the number of requests to be destroyed, only a subset of requests is removed. The subset is iteratively generated as explained below. At the first iteration, the first request of the subset is selected randomly among the requests using the spoke center and is removed. Then, at each iteration, an already removed request $i$ is randomly selected and the next request to remove, say $j$, is chosen among the requests using the spoke center with a probability increasing with the proximity of the pickup locations of $i$ and $j$. The randomization is regulated by a parameter $p\geq 1$, the bigger $p$ is, the more randomness is allowed.

\subsubsection{Late Requests Removal}\label{sec:laterem}
As already explained, during the ALNS the visit of infeasible solutions is allowed and infeasibilities, i.e. time windows and lifetime violations, are taken into account with penalty terms in the objective function. This heuristic aims at reducing the lifetime constraints violations of the current solution by removing the requests with the highest lifetime violations, if any, in order to give them the chance to be re-routed with no lifetime violation or at least a smaller one. If the solution has no late requests, this heuristic cannot be chosen by the ALNS. On the other hand, if the number of late requests of the current solution, $lr$, is smaller than the number of requests to remove, the heuristic removes $lr$ requests only.
Note that, the number of requests removed by the heuristic could not be equal to the given number of requests to remove.

\subsection{Repair Heuristics}\label{sec:lum:repair}
In the ALNS algorithm, the repairing operations are performed by seven insertion heuristics. In general, each of these heuristics takes as an input a partial solution produced after a destruction heuristic and a number of requests to insert (unplanned requests), and gives as an output a new solution in which all requests are inserted. In the ALNS, we use both repair heuristics that do not consider stabilization and heuristics in which the use of stabilization at spokes centers is evaluated for the insertion. 
The five following heuristics, denoted as {\em Best Insertion} and {\em Regret-k} (with $k=2,\ldots,5$), do not use stabilization and have been already successfully employed in the literature: 
\begin{itemize}
\item Best Insertion is a construction heuristic based on greedy criteria. At each iteration, the insertion cost of each unplanned request is calculated for each possible  position. The request with the minimum insertion cost is then inserted in its best position, and a new iteration starts. The heuristic terminates when all the unplanned requests are added to the solution \cite{ropke2006}. 
\item
The Regret-$k$ heuristic have been used for example by Potvin and Rousseau \cite{potvinrosseau} and by Ropke and Pisinger \cite{pisinger2007} for the Vehicle Routing Problem with Time Windows.
Let $R$ be the set of unplanned requests and, for each request $i \in R$, let $\Delta f_{i}^p$ be the lowest cost of inserting $i$ in the $p$-th best route at the best position. 
In the Regret-$k$ heuristic, the following procedure is iteratively performed  until all the requests are inserted. At each iteration, the request selected for insertion at its best position is $$i^*=argmax_{i \in R}\{\sum\limits_{p=1}^k (\Delta f_{i}^p-\Delta f_{i}^p)\},$$
where $\sum\limits_{p=1}^k (\Delta f_{i}^p-\Delta f_{i}^p)$ is the {\em regret value} of request $i$. 
In our algorithm, we consider Regret-$k$ heuristics with values of $k$ between 2 and 5, resulting in four different heuristics.
\end{itemize}

\medskip\medskip
The following two heuristics have been developed to evaluate the insertion of requests both with and without stabilization (at the spoke centers). These heuristics, denoted as {\em Best Insertion with spokes} and {\em Best Request-spoke Insertion}, are adaptations of repair heuristics proposed for the Pick-up and Delivery Problem with transfers \cite{masson2013adaptive, mitrovic2006pickup}.

\subsubsection{Best Insertion with spokes}
This heuristic evaluates the insertion of each unplanned request both with stabilization (i.e., by delivering it to a spoke center) and without stabilization (i.e., by delivering it directly to the HUB). In particular, for each unplanned request $i$, first the best insertion cost without the use of stabilization is calculated. Then, the best insertion with stabilization of the request is evaluated.
At this aim, the stabilization of request $i$ at a given spoke is modeled by adding two new distinct nodes: the delivery spoke node, where the request is delivered for the stabilization, and the pick-up spoke node, where the request is picked up after the stabilization. Obviously, the delivery spoke node must be visited before the pick-up spoke node.
The best insertion with stabilization is evaluated as follows. For each spoke, we first evaluate the best positions of the pair pick-up node (of the request $i$) and of the delivery spoke of $i$. Of course the two will be on the same route. Then, the best insertion for the pick-up spoke of request $i$ is computed. Note that, recalling that transfers are allowed at the spoke centers, the delivery spoke node and the pick-up spoke node could be inserted in different routes. The best insertion with stabilization is the one involving the spoke that gives the smallest overall insertion cost.
On the other hand, the best insertion without stabilization of the request $i$ is evaluated as in the Best Insertion heuristic.
Finally, the insertion costs of $i$ with or without stabilization are compared and the insertion with the smallest cost is performed.


\subsubsection{Best Request-spoke Insertion}
The Best Request-spoke Insertion heuristic inserts each unplanned request with stabilization.
In fact, even if in some cases the insertion without stabilization could be cheaper, in terms of traveled distance, the insertion of one or more spokes in the solution may help future insertion operations.

More precisely, for each unplanned request $i$, the heuristic evaluates and performs the best insertion with stabilization as in the heuristic Best Insertion with spokes. It is important to notice that the stabilization of unplanned requests can be performed in spokes already used in the solution.




\subsection{Adaptive Component and heuristic selection}
 In the ALNS algorithm,
at each iteration a pair of destroy-repair heuristics (described in sections \ref{sec:lum:destr} and \ref{sec:lum:repair})  is selected with a certain probability. 
As proposed in \cite{ropke2006}, the selection is made using the \emph{roulette wheel selection principle} described below.

Let $w_i$ be the {\em weight} of heuristic $i$. Then, at each iteration, the heuristic $j$ is chosen with probability
$$\frac{w_j}{\displaystyle \sum_{i=1}^k w_i}$$

The weights of the heuristics are automatically adjusted during the algorithm taking into account the performances of the heuristics. 
We keep track of these performances by assigning {\em scores} to the heuristics. At the end of each iteration the scores of the destroy and repair heuristics used are increased by $\sigma_1$, $\sigma_2$ or $\sigma_3$, as explained below. If the solution produced by the last heuristic pair is new and better than the best known solution, the score is increased by $\sigma_1$; if the solution produced by the last heuristic pair has not been yet accepted and better than the current solution then $\sigma_2$ is used; if the solution produced by the last heuristic pair, completely new, has not been yet accepted and is worse than the current solution $\sigma_3$ is used. A list of all the visited solutions is kept during the segment. The scores for both the destroy and repair heuristics used during an iteration are updated equally, as we cannot specify whether the destroy or the repair heuristic was the one with the good performance. In order to make a more precise search, the algorithm is divided into {\em segments} of $I=100$ iterations each. At the beginning of the first segment all the heuristics have the same weight. Then, at the end of each segment $j$, the weight of each heuristic is adjusted on the basis of the scores gained during segment $j$. At the end of each segment the scores are set to zero. More precisely, the weight that will be used for heuristic $i$ in the segment $j+1$ reads as follows, $$w_{i,j+1}=w_{ij}(1-r)+r\frac{\pi_i}{\theta_i}$$
where $\pi_i$ is the score obtained by the the $i$-th heuristic during the last segment and $\theta_i$ is the number of times the heuristic was used. Parameter $r$ controls how fast the algorithm reacts to the weight change. If $r=0$ the same weights will be used during all the algorithm, instead if $r=1$ the scores attained in the last segment will entirely determine the weights to use in the following segment.

\subsection{Acceptance Criteria} \label{sec:lum:accept}
In order not to get stuck in a local minima the acceptance criteria from \emph{simulated annealing} is used. Let $s$ be the current solution, $s'$ will be accepted with probability $e^{(f(s')-f(s))/T}$, where $T>0$ is the \emph{temperature}.
The initial temperature is $T_{0}$ and decreases at each iteration according to the expression $T=T\cdot c$ where $0<c<1$ corresponds to the \emph{cooling rate}. $T_0$ is calculated looking at the initial solution. Let $f(s_0)$ be the cost of the initial solution, as in \cite{ropke2006} the initial temperature is calculated in such a way that if the solution is $w\%$ worse than the current solution, the solution is accepted with a probability of $0.5$. More precisely the initial temperature is calculated as follows $$T_0=\frac{-w \cdot f(s_0)}{\log(0.5)}$$

\subsection{Post-Segment Procedure}\label{sec:spoke}

In the ALNS, at the end of each segment of $I=100$ iterations, a procedure is executed, called {\em Spoke Insertion} heuristic, in order to try to reduce lifetime violations of the current solution $s$, if any. The Spoke Insertion tries to stabilize some requests in order to remove or reduce  lifetime violations as follows. For each route, the last request $r_l$ violating the lifetime constraints and not already stabilized in a spoke center is detected, if any. Afterward, the spoke center $j$ that minimizes the sum of the distances to the pick-up node node of $r_l$ and to the subsequent node in the route is chosen. The spoke center is then inserted in the route right after the pickup-node of $r_l$. In practice, the insertion of $j$ in the route may generate new lifetime violations in the requests that were picked up by the vehicle before $r_l$. Hence, also such requests, if not already stabilized in a spoke, are re-routed through $j$ for stabilization. 
Once all the routes with lifetime violations have been processed, the Spoke Insertion procedure terminates and a check on the output solution is performed. If the output solution is better than the input solution (i.e., the current solution $s$) in terms of evaluation function, it is taken as a new current solution. (Otherwise, the current solution is not changed.) Furthermore, if the output solution is also better than the best known solution (best known feasible solution), the best solution (best known feasible solution) is updated.

\begin{algorithm}
\caption{ALNS Algorithm}\label{algo:lum:algo1}
{\scriptsize
\begin{tabbing}
{\bf{Generate}} Initial Solution, $s_0$\\
$s_{best}\leftarrow s_0$\\
$s_{curr}\leftarrow s_0$\\
{\bf{Repeat}} \\
\quad {\bf{Selection}} \emph{Destroy} and \emph{Repair} heuristics \\
\quad $s\leftarrow s_{cur}$\\
\quad $s\leftarrow Destroy(s)$ \\
\quad $s\leftarrow Repair(s)$ \\
\quad {\bf{If}} $f(s)<f(s_{best})$ {\bf{Then}}  $s_{best}\leftarrow s$; $s_{cur}\leftarrow s$\\
\quad {\bf{Else}}  {\bf{If}}  $Accepted(s, s_{cur})$ {\bf{Then}}  $s_{cur}\leftarrow s$\\
\quad {\bf{If}} end of segment {\bf{Then}} \\
\quad \quad $s_{sp} \leftarrow SpokeInsertion(s_{cur}) $ (See Section \ref{sec:spoke})\\
\quad\quad {\bf{If}} $f(s_{sp})<f(s_{cur})$ {\bf{Then}} $s_{cur}\leftarrow s_{sp}$ \\
\quad\quad {\bf{If}} $f(s_{sp})<f(s_{best})$ {\bf{Then}}  $s_{best}\leftarrow s_{sp}$ \\
{\bf {Until}} a maximum number of iterations is reached
\end{tabbing}
}
\end{algorithm}


\subsection{Initial Solution} \label{sec:lum:init}

In the proposed ALNS algortihm,  Large Neighborhood Search (LNS) procedures have been used to produce an initial solution with small lifetime violations. In fact, for the problem under study, we observed that a too high total lifetime violation of the initial solution may negatively affect the performances of the overall algorithm. 
At this aim, different starting solutions are generated by employing each of the repairing heuristics described in Section \ref{sec:lum:repair}. In this case, the repair heuristics are used as constructive heuristics for routing all the requests to the vehicles.
Then, a 5-iterations LNS is applied to each starting solution generated as described above, with the aim of reducing the lifetime violations, if any. More precisely, at each iteration of the LNS procedures, the destruction heuristic Late Requests Removal (see Section \ref{sec:laterem}) is first applied to remove late requests. Then, the solution is repaired with an heuristic randomly chosen among the pool of repairing heuristics and finally the post processing procedure Spoke Insertion (introduced in Section \ref{sec:spoke}) is applied. At the end of each iteration, the new generated solution is kept only if it is better than the previous solution. 
In a preliminary experimental campaign, it has been observed that these small LNS procedures are able to sensibly reduce the lifetime violations of the starting solutions. Obviously, the LNS procedures are not applied to starting solutions with no lifetime violation.
After applying the LNS to each starting solution with lifetime violations, the best solution obtained so far is chosen as initial solution of the ALNS algorithm.

\section{Real-world data and instance description}
\label{sec:LUM:inst}

The MILP formulations and the ALNS algorithm have been tested on instances generated from real data, arising from the case study of the Local Healthcare Authority of Bologna, Italy. 
In Bologna, a reorganization plan for the territorial network of the Bologna Analysis Laboratories was conducted. The reorganization consisted in the activation of a Single Metropolitan Laboratory (LUM) for the analysis of  the all biological samples collected in the metropolitan area of Bologna.
In the area, there are 46 (blood and biological samples) draw centers geographically distributed. The opening days and hours of each center depend on the center itself and range from 7 am to 10 am, from Monday to Saturday. According to the real data, in each center, a sample is drew about every 3 minutes. Table \ref{tab:lum:timetable} reports the number of draw centers opened each day and the average opening hours of the centers in minutes. The lifetime of each sample is of 120 minutes. Hence, from its withdrawal, a sample must be delivered to the main hospital or to a spoke center within $120$ minutes. The fleet of vehicles that can be used for transporting samples is composed of 26 vehicles for weekdays and of 16 for Saturday. Vehicles are located in 8 different depots. Each day, the stabilization process can be performed in 12 spoke centers and requires about $30$ minutes. The extra lifetime gained by a stabilized sample is of $90$ minutes (starting from the end of the stabilization process). The service time required by a vehicle to load or unload samples at a draw center or a spoke is about 10 minutes. 
\begin{table}
\begin{center}
\scriptsize
\begin{tabular}{l r r r r r r }
& \textbf{ Monday}& \textbf{ Tuesday} &\textbf{ Wednesday}&\textbf{ Thursday}&\textbf{ Friday}&\textbf{ Saturday}\\
\hline
Draw centers& 31& 33& 30&36&30&16\\
Av. Time (min) & 100.65	&100.00&	100.00& 	99.17 &	100.00 &	108.75  \\ \hline
\end{tabular}
\end{center}
\caption{Number of opened draw centers and average opening hours (in minutes).}
\label{tab:lum:timetable}
\end{table}

\begin{figure}[H]
\begin{subfigure}{.5\linewidth}
\centering
\includegraphics[width=0.9\textwidth ]{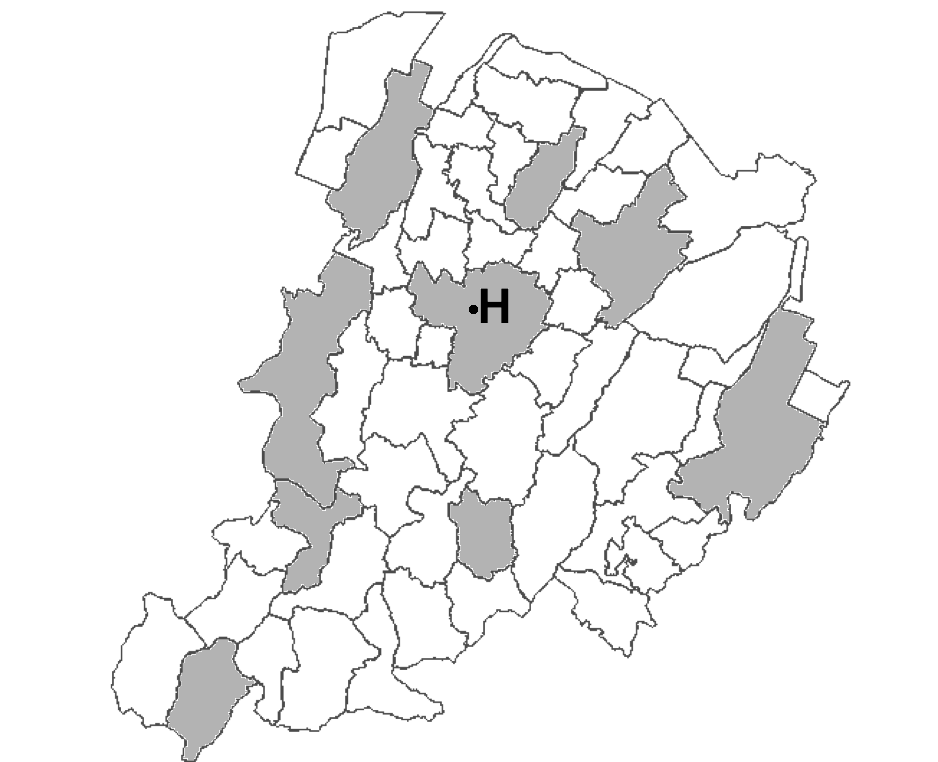}
\caption{spoke Centers}
\label{fig:sub1}
\end{subfigure}%
\begin{subfigure}{.5\linewidth}
\centering
\includegraphics[width=0.9\textwidth ]{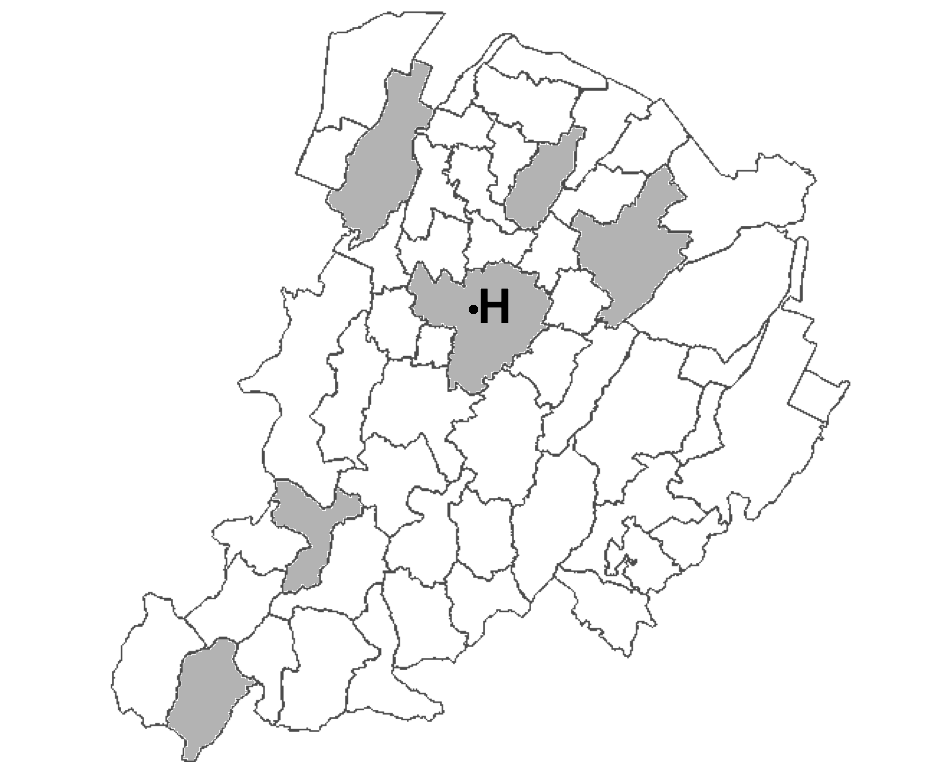}
\caption{Depots}
\label{fig:sub2}
\end{subfigure}\\[0ex]
\begin{center}
\begin{subfigure}{.5\linewidth}
\centering
\includegraphics[width=0.9\textwidth]{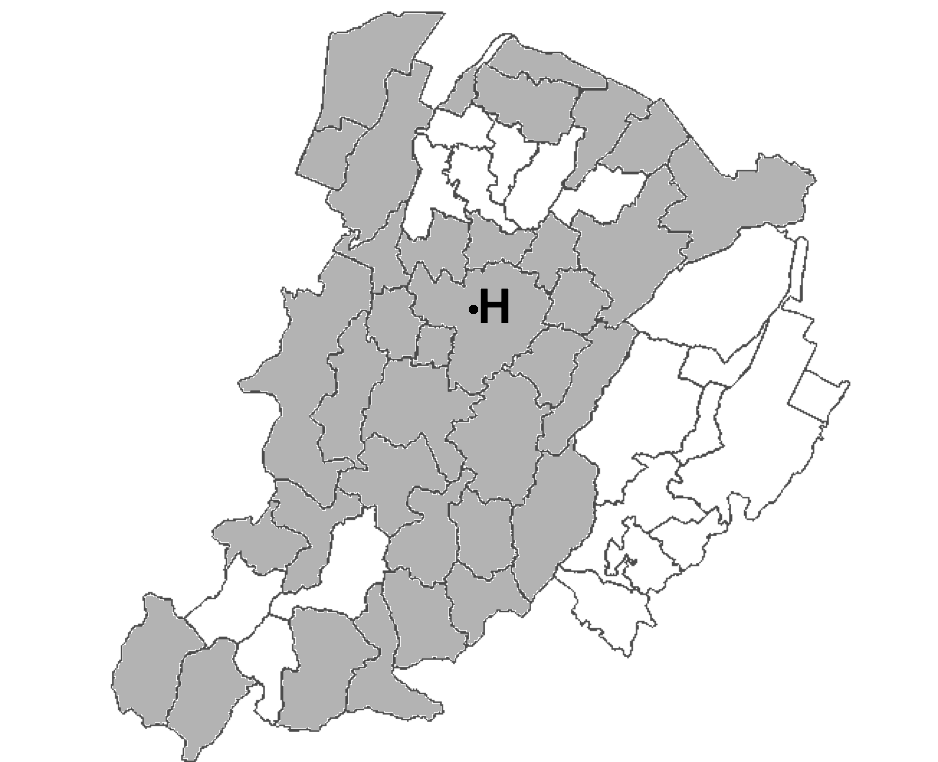}
\caption{Draw centers}
\label{fig:sub3}
\end{subfigure}
\end{center}
\caption{Locations of draw centers, spoke Centers and Depots.}
\label{lum:mappe}
\end{figure}

Figure \ref{lum:mappe} reports the geographical distribution of the draw centers, spokes and depots in the metropolitan area of Bologna. More precisely, each  area highlighted in grey contains at least one of the above mentioned centers. The location of the main hospital (HUB), denoted by  $H$, is also shown. 

From the real case described above, different sets of small and large instances have been generated. 
More precisely, ten small instances have been produced in order to compare the performances of  the two MILP formulations introduced in Section \ref{sec:milp} and the ALNS algorithm. Furthermore, 32 large instances have been generated for solving the real problem.

In the small instances, the number of requests (i.e., samples) and vehicles range from 5 to 20 and from 3 to 10, respectively, while the number of spokes is one or two. The details of the ten small instances are reported in Table \ref{tab:small}. More precisely, in the table, Column 2 reports the time step used in the two time-indexed formulations $MILP_1$ and $MILP_2$ (i.e., one or five minutes). Obviously, the bigger the time step is, the smaller the number of variables of the formulations is. Columns 3--5 respectively report the number of requests $|R|$, vehicles $|K|$ and spokes $|SP|$ of the instances.
These instances have been designed to also test the different transportation modes of the samples (i.e., directly to the main hospital, or first to a spoke and then to the hospital).
For example, in Instance 2 a feasible solution can be obtained only by using stabilization at the spokes, while in Instance 7 an optimal solution exists that does not require sample stabilization.
\begin{table}
\begin{center}
\scriptsize
\begin{tabular}{l| r r r r }
Id &Time step &$|R|$&$|K|$&$|S|$\\
\hline
1&1&5&3&1\\
2&1&8&3&1\\
3&1&9&4&1\\
4&5&9&4&1\\
5&5&12&6&1\\
6&5&12&6&2\\
7&5&12&7&2\\
8&5&15&8&1\\
9&5&15&9&2\\
10&5&20&10&2\\
\hline
\end{tabular}
\end{center}
\caption{Characteristics of the small instances.}
\label{tab:small}
\end{table}

\medskip

In the large instances, all vehicles and all draw and spoke centers have been included. Furthermore, since the daily number of samples to manage is extremely high (i.e., more than 500 and 1000 on Saturday and weekdays, respectively),  samples have been grouped in batches according to different time spans. In fact, recalling that a biological sample is produced every 3 minutes at each draw center, samples have been grouped in batches every either $30$, $45$ or $60$ minutes of activity of a draw center. Hence, as an example, when samples are grouped in batches using a time span of $30$ minutes, a batch, i.e. a transportation request, contains about 10 samples.
The earliest limit of the time window assigned to each batch is set equal to the ending production time of the last sample of the batch, i.e., the production time the last sample of the batch, and the latest limit of the time window is always set to $30$ minutes after the closing time of the draw center. Thus, the time window width varies depending on the batch. The lifetime  assigned to each batch is set to 120 minutes after the production time of the oldest sample of the batch, i.e., the sample of the batch which is generated first. In this way we ensure that, if a batch is delivered on time to the main hospital, every sample contained in the batch is delivered on time, too. As a consequence, the bigger the time span used for grouping the samples is, the shorter the lifetimes of the batches are, since the cardinality of each batch is bigger.

In general, when grouping samples of a draw center in batches using a given time span, the number of batches generated may vary according to the opening hours of the center and, if the time span and the opening hours are not divisible numbers, to the rounding procedure used. 
More precisely, let $ts$ be the time span used and $o_k$ be the opening hours (in minutes) of center $k$. The batches generated may be $\lceil o_k/ts \rceil $ or $\lfloor o_k/ts \rfloor $. In the first case (second case), the last generated batch of each draw center may contain a smaller (bigger) number of samples, resulting in a batch with a longer (shorter) lifetime.
According to the above criteria, for each time span $ts$, we generate two sets of large instances, denoted as Set $A$ and Set $B$,  containing  $\lceil o_k/ts \rceil$  and   $\lfloor o_k/ts \rfloor $ batches for each center $k$, respectively. 
As an example, let us assume that the opening hours of the draw center $k$ are $o_k=300$ minutes and that $ts=45$ minutes. Hence, $\lceil o_k/ts \rceil=7$ and $\lfloor o_k/ts \rfloor =6$. 
In the first case, we get 7 batches: the first 6 batches with 15 samples and lifetime of 78 minutes and the last batch with 25 samples and a shorter lifetime of 48 minutes.
In the second case, we get 6 batches: the first 5 batches contain 15 samples and lifetime 78 minutes, and the last batch contains 10 samples and has a longer lifetime of 93 minutes.

Tables \ref{tab:lum:reqceil} and \ref{tab:lum:reqfloor}  show the number of requests/batches in the instances of Set $A$ and Set $B$, generated by grouping samples according to the two different rounding procedures, with a time span of  $60$, $45$ and $30$ minutes respectively, from Monday to Saturday. Hence, as detailed above, instances of Set $A$ generally contain a bigger number of batches than the instances of Set $B$, but possibly with longer lifetimes.
Note that, in each set, since the opening hours of the draw centers depend on the day of the week, the number of requests for each day varies.
Furthermore, when a time span of 30 minutes is considered, the number of requests in the instances of Set $A$ and Set $B$ related to Monday, Wednesday, Friday and Saturday are the same, resulting in exactly the same instances. As a consequence, the number of different large instances obtained so far is 32.

\begin{table}[H]
\begin{center}
\scriptsize
\begin{tabular}{l c| r  r r r r r}
\textbf{Set} & \textbf{Time Span} & \textbf{ Monday}& \textbf{ Tuesday} &\textbf{ Wednesday}&\textbf{ Thursday}&\textbf{ Friday}&\textbf{ Saturday}\\
\hline
$A_1$&60  &  61 & 65 &60 &71 &59 &34\\
$A_2$&45  &  76 & 81 &  72 & 87 & 73 & 42\\
$A_3$&30  &104&111&100&120& 100 & 58\\
\hline
\end{tabular}
\end{center}
\caption{Number of requests/batches per time span  and day in instances of Set $A$.}
\label{tab:lum:reqceil}
\end{table}
\begin{table}[H]
\begin{center}
\scriptsize
\begin{tabular}{l c| r  r r r r r}
\textbf{Set} & \textbf{Time Span} & \textbf{ Monday}& \textbf{ Tuesday} &\textbf{ Wednesday}&\textbf{ Thursday}&\textbf{ Friday}&\textbf{ Saturday}\\
\hline
$B_1$&60  &  43 & 45 &40 &48 &41 &24\\
$B_2$&45  &  62 & 66 &  61 & 72 & 60 & 35\\
$B_3$&30  &104&109&100&118& 100 & 58\\
\hline
\end{tabular}
\end{center}
\caption{Number of requests/batches per time span  and day in instances of Set $B$.}
\label{tab:lum:reqfloor}
\end{table}

\section{ Experimental Results }
\label{sec:LUM:res}
In this section, the experimental results on the instances described in the previous section are presented. All tests have been performed on a PC equipped with Intel i5 processor and 8 Gb of RAM. This section is organized as follows. In Section \ref{sec:LUM:inst} a tuning experimental campaign devoted to properly set some parameters of the ANLS algorithm is presented. In Sections \ref{sec:LUM:milp_res} and \ref{sec:LUM:anls_res}
, the computational results on small and large instances are respectively reported.

\subsection{ ALNS Tuning}
\label{sec:LUM:tune}

In this section, the results of an experimental tuning campaign carried out to properly set some components and parameters of the ALNS algorithm are reported.

%


On the basis of a preliminary experimental analysis, it follows that the parameters that highly affect the performance of the ALNS are $r$ and the percentage of requests to remove and re-insert at each iteration, denoted as $rem$ in what follows. Recall that $r$ controls how the weights assigned to each heuristic change during the algorithm: the smaller $r$ is the more importance is given to the whole performance history of the heuristics during the algorithm when updating their weights. 
As a consequence, in the tuning phase, different values of $r$ and $rem$ have been evaluated, while the other parameters have been set as proposed in \cite{ropke2006}, i.e., $\sigma_1=33$, $\sigma_2=9$, $\sigma_3=19$, $C_{rate}=0.99975$ and $w=0.05$. 
Furthermore, in the tuning phase, different variants of the insertion heuristics have been tested, which take into account, with different accuracy levels, how the insertion of a request in a given route affect other routes of the solution.
In fact, in the problem under study, the requests can be transferred from one vehicle to another at the spoke centers, implying that routes may affect each other. As an example, let us suppose that request $i$ is delivered by vehicle $k$ to a spoke $sp$ and is picked up at $sp$ by a vehicle $q$ after the stabilization. Then, a change in the route $k$ may lead to request $i$ ending the stabilization process at spoke $sp$ earlier or later than before, implying a possible timing change in the route $q$. 
In practice, taking into account the changes on the affected routes during the insertion of requests performed by an insertion heuristic increases the accuracy of local insertion decisions, but, at the same time, increases the complexity of the heuristics. 
In order to evaluate how the global performance of the ALNS algorithm is influenced by the evaluation of the affected routes in the insertion heuristics, the following three configurations have been tested:
\begin{itemize}
    \item Configuration 1 ($C_1$): The affected routes are never taken into account during the insertion evaluation. Hence,   heuristics evaluate the insertion of a request in a route $k$  considering route $k$ only.
    \item Configuration 2 ($C_2$): The affected routes are always taken into account during the insertion evaluation. Hence, a heuristic evaluates the insertion of a request in a route taking into account how that insertion affect other routes.
    \item Configuration 3 ($C_3$): The affected routes are taken into account only during the evaluation of an insertion  of a request together with a spoke center. Hence, only the insertion heuristics Best Insertion with spokes and Best Request-spoke Insertion consider the affected routes during the insertion evaluation. 
    This configuration arises from the fact that inserting a request together with a spoke center could have a higher impact on other routes than the insertion of a single request.
\end{itemize}

In the tuning phase, the  three configurations introduced above and  different values for parameters $r$ and $rem$ have been tested. More precisely, we consider $r\in\{0.1, 0.5, 0.8\}$ and $rem\in\{0.01, 0.05, 0.1, 0.15\}$. The values of $r$ have been chosen in order to test the algorithm with a high, medium, or small sensitivity to the weights' change. The values of $rem$ came out from a preliminary experimental campaign, which showed that the proposed algorithm performs better, with respect to solution quality and computational time, when the  percentage of requests to be removed/inserted at each iteration is quite small.
All combinations of the values of $r$ and $rem$ and the three configurations have been tested on 4 real-world instances, i.e., Saturday and Friday of Set $A_3$, and Wednesday and Thursday of set $A_2$.
Hence, 144 experiments have been performed in total: each configuration has been tested on 48 experiments (one for each instance, $r$ and $rem$ values), while each value of $r$ and $rem$ has been tested on 48 and 36 experiments, respectively.
The main indicators used to evaluate the performances of the ALNS algorithm are the percentage of feasible solutions found, the  solution values, and the computation time. 

 The results of the tuning campaign are summarized in Tables \ref{lum:tuning30} and \ref{lum:tuning45}.
 On the smallest instance, Saturday of Set $A_3$ (Table \ref{lum:tuning30}), the ALNS was able to find a feasible solution in all the runs. The best solution is found with configuration $C_2$ , $r=0.1$ and $rem=0.01$.
 As Table \ref{lum:tuning30} shows, on Friday of Set $A_3$, the algorithm is  able to find a feasible solution with $r=0.1$, only. The value of $rem$ that provides the highest percentage of feasible solutions is 0.05 and the best solution is found with configuration $C_1$, $r=0.1$ and $rem=0.05$. 
 As shown in Table \ref{lum:tuning45}, on the instance Wednesday of Set $A_2$, the ALNS with configuration $C_1$ is able to find a feasible solutions in the $92\%$ of the runs, against the $50\%$ and $67\%$ of configuration $C_2$ and configuration $C_3$, respectively. The values $r=0.1$ and $rem=0.05$ are the best ones in terms of feasibility ($100\%$ and $89\%$, respectively). The best solution is found with configuration $C_1$, $r=0.1$ and $rem=0.01$, but $rem=0.01$ has only a $44\%$ feasibility. The second best solution (with a gap of less than $2\%$ from the best solution) is found with configuration $C_1$, $r=0.1$ and $rem=0.05$ (in this case, ALNS is able to find a feasible solution in the $89\%$ of the runs).
 On Thursday of Set $A_2$ (Table \ref{lum:tuning45}), configuration $C_1$ is the only one enabling the ALNS to find feasible solutions and the best solution is found with $r=0.1$ and $rem=0.05$.
 
 To conclude, configuration $C_1$ performs better than the other configurations in all the instances except Saturday of Set $A_3$. However, as already stated, this is the smallest and easiest instance, on which all the configurations perform reasonably well. The other instances are much bigger and, as shown in the Tables \ref{lum:tuning30}-\ref{lum:tuning45}, configuration $C_1$ provides the best results, in terms of percentage of feasible solutions, as well as best solutions. 
 Hence, configuration $C_1$ is the more reliable and effective of the three.
 Furthermore, the ALNS algorithm with this configuration always finds the best solution with $r=0.1$ and $rem=0.05$, except for Wednesday of Set $A_2$ where, as mentioned before, the best solution is found with $rem=0.01$ (even if the second best solution is found with $rem=0.05$). The values $r=0.1$ and $rem=0.05$ seem in general to be the most effective.
 Regarding computational times, a comparison with last rows of Tables \ref{lum:tuning30} and \ref{lum:tuning45} shows that the combination configuration $C_1$, $r=0.1$ and $rem=0.05$ require in general higher computational times than the other combinations on the four instances, on average. Anyway, such a higher computational effort is compensated by better performances in terms of solution feasibility and quality.


\begin{table}
\scriptsize
\begin{center}
\begin{tabular}{| c c || c   c | c  c | c  c || c   c | c  c | c  c |}
\hline
\multicolumn{2}{|c|}{} &\multicolumn{6}{|c|}{Saturday $A_3$}&\multicolumn{6}{|c|}{Friday $A_3$}\\
\hline
\multicolumn{2}{|c|}{} & \multicolumn{2}{|c|}{ $C_1$ }         & \multicolumn{2}{|c|}{$C_2$ } & \multicolumn{2}{|c|}{ $C_3$  }    & \multicolumn{2}{|c|}{ $C_1$ }         & \multicolumn{2}{|c|}{$C_2$ } & \multicolumn{2}{|c|}{ $C_3$  }   \\ \hline
r&rem&Solution&Time&Solution&Time&Solution&Time&Solution&Time&Solution&Time&Solution&Time\\\hline \hline
0.1&0.01&1407&21.44&1197&26.75&1295&26.16&2119&91.74&-&87.04&2202&99.63\\
0.1&0.05&1373&115.67&1361&153.23&1257&44.14&2104&524.36&2515&624.33&2264&750.54\\
0.1&0.1&1385&158.08&1287&68.93&1458&36.57&2164&469.08&-&227.69&2467&578.68\\
0.1&0.15&1603&305.53&1427&63.12&1648&30.43&2307&242.72&-&343.71&-&368.11\\
0.5&0.01&1419&23.90&1290&27.83&1212&25.67&-&135.78&-&149.92&-&154.15\\
0.5&0.05&1411&76.51&1359&93.99&1398&126.22&-&259.72&-&170.07&-&319.35\\
0.5&0.1&1611&165.35&1469&102.84&1647&144.02&-&166.76&-&282.59&-&196.65\\
0.5&0.15&1620&146.17&1726&54.20&1620&143.51&-&112.05&-&230.16&-&259.02\\
0.8&0.01&1483&83.09&1245&30.71&1272&49.75&-&201.43&-&208.25&-&152.52\\
0.8&0.05&1488&46.36&1518&114.24&1540&163.90&-&150.68&-&229.12&-&156.79\\
0.8&0.1&1613&105.53&1652&129.92&1551&154.51&-&141.63&-&144.48&-&171.87\\
0.8&0.15&1726&77.94&1617&61.43&1620&132.14&-&122.39&-&149.81&-&164.36\\
 \hline
&Av.&1511.58&110.46&1429.00&77.27&1459.83&89.75&2173.5&218.19&2515&237.26&2311&280.97\\
\hline
\end{tabular}
\end{center}
\caption{Tuning results on instances Saturday and Friday of Set $A_3$.}
\label{lum:tuning30}
\end{table}

\begin{table}
\scriptsize
\begin{center}
\begin{tabular}{| c c || c   c | c  c | c  c || c   c | c  c | c  c |}
\hline
\multicolumn{2}{|c|}{} &\multicolumn{6}{|c|}{Wednesday $A_2$}&\multicolumn{6}{|c|}{Thursday $A_2$}\\
\hline
\multicolumn{2}{|c|}{} & \multicolumn{2}{|c|}{ $C_1$ }         & \multicolumn{2}{|c|}{$C_2$ } & \multicolumn{2}{|c|}{ $C_3$  }    & \multicolumn{2}{|c|}{ $C_1$ }         & \multicolumn{2}{|c|}{$C_2$ } & \multicolumn{2}{|c|}{ $C_3$  }   \\ \hline
r&rem&Solution&Time&Solution&Time&Solution&Time&Solution&Time&Solution&Time&Solution&Time\\\hline \hline

0.1&0.01&2070&57.78&2287&176.55&2176&105.88&-&159.60&-&80.22&-&151.74\\
0.1&0.05&2112&282.43&2140&198.26&2389&291.97&2266&438.53&-&491.10&-&605.23\\
0.1&0.1&2232&154.49&2284&270.73&2168&161.72&-&287.70&-&300.05&-&434.66\\
0.1&0.15&2460&226.36&2147&405.24&2329&101.68&2572&886.82&-&232.78&-&357.48\\
0.5&0.01&2323&81.39&-&127.66&-&216.28&-&130.06&-&141.06&-&144.33\\
0.5&0.05&2438&94.50&2492&202.37&2357&266.22&2452&236.93&-&199.60&-&201.33\\
0.5&0.1&2477&109.71&-&180.20&2714&114.26&-&160.73&-&233.93&-&133.22\\
0.5&0.15&2858&122.45&-&98.55&2439&139.60&-&149.20&-&159.99&-&158.89\\
0.8&0.01&-&95.59&-&149.27&-&151.17&-&112.85&-&113.75&-&136.24\\
0.8&0.05&2268&155.33&2578&141.01&-&102.35&-&153.79&-&113.54&-&152.87\\
0.8&0.1&2579&85.85&-&124.97&2545&11.12&-&202.43&-&164.85&-&158.83\\
0.8&0.15&2406&106.31&-&108.02&-&110.05&-&140.77&-&122.77&-&121.76\\\hline
&Av.&2383.91&131.01&2321.33&181.90&2389.63&147.69&2430.00&254.95&-&196.14&-&229.72\\
\hline
\end{tabular}
\end{center}
\caption{Tuning results on instances Wednesday and Thursday of set $A_2$.}
\label{lum:tuning45}
\end{table}

%


\subsection{Results on small instances}
\label{sec:LUM:milp_res}

In order to test the two MILP formulations of the problem and the effectiveness of the ALNS algorithm, the ten small instances of different sizes  generated from real-life instances and introduced in Section \ref{sec:LUM:inst} have been used. As already stated, these instances have been generated in order to test the formulations on different aspects of the problem, e.g., sample stabilization,  multiple visits of the vehicles at spoke centers. The two mathematical formulations $MILP_1$ and $MILP_2$ have been solved by Gurobi 9.0 on a 3.2 GHz computer with 12 cores, 64 Gb of RAM with a time limit of 16 hours. 
The ALNS algorithm has been executed on a 2.5 GHz Quad-core processor and 16 Gb of RAM, using the parameter combination selected in the tuning phase presented in Section \ref{sec:LUM:tune}, i.e., Configuration $C_1$, $rem=0.05$ and $r=0.1$, and setting $It_{max}=10,000$. 

Table \ref{tab:res_small} reports a summary of the results of $MILP_1$, $MILP_2$ and of the ALNS algorithm.
In the table, for each instance, Columns 2--4 and 5--7 respectively report the results related to $MILP_1$ and $MILP_2$. More precisely, ``LB'' is the best lower bound found at the root node, ``Sol'' is the value of the best solution and "time" is the computation time in seconds. 
A ``*'' indicates that Gurobi was not able to certify the optimality of the solution within the time limit.
Finally, Columns 8--10 show the results of the ALNS algorithm, in which, for each instance, ``Av." is the average solution found over 5 runs of the algorithm, ``\# Opt." is the number of times the optimal solution is found out of the five runs and ``time'' is the average computation time in seconds.

\begin{table}
\begin{center}
\scriptsize
\begin{tabular}{|l |r r r| r r r |r r r| }
\hline
 & \multicolumn{3}{|c|}{ $MILP_1$ }  & \multicolumn{3}{|c|}{$MILP_2$}          & \multicolumn{3}{|c|}{ALNS }  \\ \hline
Id & LB & Sol &time&LB & Sol & time & Av.&  \# Opt. &time\\
\hline
1&227&227&601.53&184&227&526.83&227.00&5&0.84\\
2&366&366&1.024.73&322&366&1,171.99&366.00&5&1.15\\
3&385&385&1,862.24&385&385&1,432.54&385.00&5&0.97\\
4&328&328&19.02&260&328&19.14&328.00&5&1.34\\
5&572&572&39.11&420&572&55.86&580.80&3&1.83\\
6&572&572&187.80&311&572&57600*&583.20&3&2.11\\
7&572&572&231.62&322&572&57600*&577.60&4&2.63\\
8&569&569&87.03&569&569&90.20&570.60&3&3.11\\
9&528&528&604.88&390&638&57600*&534.60&3&4.23\\
10&302&302&1,463.58&211&302&42,088.83&358.20&0&3.80\\
\hline
\end{tabular}
\end{center}
\caption{Results on the small instances.}
\label{tab:res_small}
\end{table}

The results show that $MILP_1$ is able to find the optimal solution for all the small instances considered. In most of the cases the optimality is certified at the root node. The computation time required by  $MILP_1$ generally depends on the instance dimension, and ranges from a few dozens of seconds to about half an hour. It also seems to depend on other characteristics of the instances. In fact, when the optimal solution requires  multiple visits of a vehicle at a spoke center, the computation time seems to be higher. Also, the number of spokes seems to affect the computation time more than increasing the number of requests or vehicles.
$MILP_2$ performs similarly on the 5 smallest instances but, in general, as the size of the instances increases, it starts to struggle in finding the optimal solution. In three of the ten instances, $MILP_2$ was not able to certify the optimal solution within the time limit. In two of these cases the best solution found is actually the optimal solution, but the optimality gap was not filled up. Such a behavior could be mainly due to the strength of the linear relaxation: In fact,  the best lower bound found at root node of $MILP_1$ always equals the optimal solution value, while the one provided by $MILP_2$ is generally smaller.
As for the ALNS, it has proven to be quite effective when compared to the MILP models.  Note that, in Column ``\#Opt.'', in four instances the optimal solution was found in each run, while in the other instances it was found only in some of the 5 runs, as indicated. Instance 10 is the only one where the ALNS was not able to find the optimal solution in the 5 runs.
However, even the not optimal solutions are in general very close to the optimum: on Instances 5-9, there is an average gap of $1.92\%$ between the optimal solution and the ALNS average solution while on instance 10 the gap is about $15\%$. As for the computational times, the ALNS is obviously much faster with an average time going from 0.8 to 4.2 seconds.

\subsection{ALNS results on real-life instances}
\label{sec:LUM:anls_res}
In this section, experimental results of the ALNS algorithm on the 32 distinct large instances of $Set A$ and $Set B$ are presented. Even on the smallest instances of these sets, we point out that Gurobi running on $MILP_1$  and $MILP_2$ was not even able to find a feasible solution in hours of computation.

Tables \ref{lum:resA} and  \ref{lum:resB}
report the results on the two sets. We recall that the instances of Monday, Wednesday, Friday and Saturday of Sets $A_3$ and $B_3$ are the same.

On each instance, five runs of the algorithm have been performed employing the parameters and the algorithm configuration selected from tuning phase presented in Section \ref{sec:LUM:tune}, i.e., configuration $C_1$, $rem=0.05$ and $r=0.1$, and with $It_{max}=20,000$. As shown in Section \ref{sec:LUM:tune}, on selected instances, such parameters allow on average to find feasible solutions with small costs.

The objective of the experimental campaign on the large instances is twofold. From one side, the ability of the developed ALNS algorithm to tackle with big real-life instances can be assessed. From the other side, it allows to evaluate the best policy for batching samples in such a way that all samples can be delivered on time.  Recall that, instances of  Set $ A$ and Set $ B$ present different characteristics: instances in  Set $A$ usually have a bigger number of requests but with longer lifetimes, while instances in Set $ B$  are smaller but with shorter lifetimes of the requests. In Tables \ref{lum:resA} and \ref{lum:resB}, Columns 2--5, Columns 6--9 and Columns 10--13 show the results for sets $A_1/B_1$,  $A_2/B_2$ and $A_3/B_3$, respectively, from Monday to Saturday.  In the tables,  ``Av." and ``Best." are the average travel distance and the best travel distance, respectively, of the solutions found on the 5 runs of the algorithm,  ``Feas." is  the number of times, out of the 5 runs, a feasible solution was found, and ``time" is the average computational time in seconds. In the tables, a ``-" in  ``Av." and ``Best." means that no feasible solution has been found in the five runs.

First of all, note that the ALNS algorithm is not able to find feasible solutions for all instances of Set $B_1$, but Saturday (for which it founds the best feasible solution for that day). In fact, instances in Set $B_1$ contain requests with the shortest lifetimes and, hence, are the most {\em constrained} in terms of lifetimes. In all other instances, except for the instances related to Thursday in $A_3$ and $B_3$, the ALNS is able to find feasible solutions.
Regarding the batching policy, the results show that bigger the time spans for batching samples are, better the quality of the solutions are, in terms of objective function. As shown in Tables \ref{lum:resA} and \ref{lum:resB}, for each day, the average and the best costs of the solutions usually increase as the time span increases, i.e., from instances of Set $A_3$ (Set $B_3$) to instances of Set $A_1$ (Set $B_1$).
Nevertheless, note that by employing the rounding procedure used to generate Set $B$ and a time span of 60 minutes, the instances become too much constrained in terms of lifetimes so that a feasible solution could not exist.
The results show that using a time span of 45 minutes is the best batching policy with respect to feasibility. In fact, the ALNS always find a feasible solution in both Set $A_2$ and Set $B_2$.

In general, on all the days, the best results in terms of best and average objective function values and computational time are attained on instances of Set $A_1$.

For the single days, the best results are obtained for the instances of Set $A_1$, except for Saturday for which the best results are attained in Set $B_1$. Note that, the instance Saturday in Set $B_1$ is the smallest instance in terms of number of requests.

Regarding the biggest instances of Sets $A_3$ and $B_3$, we observe that the ALNS algorithm fails in finding feasible solutions for the instances of Thursdays and few feasible solutions are founds on Tuesdays' instances. This fact could mainly due to the big sizes of these instances, with a number of requests ranging from 109 to 120. On the other hand, especially on instances of Set $A_3$, ALNS is able to find quasi-feasible solutions with  small total lifetime violations (of about 22 minutes) and total time-windows violations (of about 1 minute), on average.

\begin{table}
\scriptsize
\begin{center}

\begin{tabular}{| l | c r c  c | c  r  c c | c r  c c |}
\hline
$Set A$ & \multicolumn{4}{|c|}{ $A_1$ ($ts=$ 60) }         & \multicolumn{4}{|c|}{$A_2$ ($ts=$ 45)}          & \multicolumn{4}{|c|}{$A_3$ ($ts=$ 30) }     \\ \hline
&Av.&Feas.&Best&time & Av.&Feas.&Best&time & Av.&Feas.&Best&time \\
 \hline
Monday&1893&5&1836&406.80&2043.6&5&2002&414.58&2320.75&4&2238&428.68\\
Tuesday&2224.2&5&2071&454.26&2409.8&5&2298&597.04&2603.5&2&2569&725.87\\
Wednesday&2040.4&5&1914&271.35&2179&5&2084&228.40&2490.75&4&2241&400.35\\
Thursday&2318.3&4&2234&436.50&2539.6&5&2,438&686.72&-&0&-&892.48\\
Friday&1777.4&5&1697&220.12&2000.6&5&1970&279.06&2144.75&4&2102&475.21\\
Saturday&1204.6&5&1180&34.82&1399.4&5&1344&34.60&1201&5&1057&105.99\\
 \hline
 Av.&1909.65&4.83&1822.00&303.97&2095.33&5.00&2022.67&373.40&2152.15&3.17&2041.40&504.76\\
 \hline
\end{tabular}
\end{center}
\caption{ALNS results on the instances of $Set A$.}
\label{lum:resA}
\end{table}

\begin{table}
\scriptsize
\begin{center}

\begin{tabular}{| l | c r c  c | c  r  c c | c r  c c |}
\hline
$Set B$ & \multicolumn{4}{|c|}{ $B_1$ ($ts=$ 60) }         & \multicolumn{4}{|c|}{$B_2$ ($ts=$ 45)}          & \multicolumn{4}{|c|}{ $B_3$ ($ts=$ 30) }     \\ \hline
&Av.&Feas.&Best&time & Av.&Feas.&Best&time & Av.&Feas.&Best&time \\
 \hline
Monday&-&0&-&269.11&1995&5&1842&334.23&2320.75&4&2238&428.68\\
Tuesday&-&0&-&283.61&2352&5&2254&427.83&2709&1&2709&576.05\\
Wednesday&-&0&-& &2060&5&1970&312.31&2490.75&4&2241&400.35\\
Thursday&-&0&-& &2380.2&5&2,286&540.33&-&0&-& \\
Friday&-&0&-& &1908&5&1770&162.12&2144.75&4&2102&475.21\\
Saturday&859.5&4&817&54.11&1058.2&5&849&57.38&1201&5&1057&105.99\\
 \hline
 Av.&-&-&-&-&1958.90&5.00&1828.50&305.70&2173.25&3.00&2069.40&397.26\\
 \hline
\end{tabular}
\end{center}
\caption{ALNS results on the instances of $Set B$.}
\label{lum:resB}
\end{table}

Summarizing, the ALNS algorithm is able to deliver all samples on time by grouping samples with a time span $ts$ of 45 minutes. This is an important and interesting result since in the real case related to the Local Healthcare Authority of  Bologna about the 40\% of samples is delivered late.

\section{Conclusion}
\label{sec:LUM:conclusion}
In this paper, a Vehicle Routing problem arising from a real-world healthcare application has been presented, concerning the transportation of biological samples from draw centers to a main laboratory. The problem is modeled as a Vehicle Routing Problem with time-window and lifetime constraints, transfers and multiple possible visits at specific nodes (i.e., the spokes), where samples can be stabilized to gain extra lifetimes. 
 Two Mixed Integer Linear Programming formulations and an Adaptive Large Neighborhood Search algorithm have been proposed for the problem. Computational experiments on different sets of instances based on real-life data provided by the Local Healthcare Authority of Bologna, Italy, are presented. A comparison between the solutions obtained with the formulations and the ALNS algorithm on small instances shows the effectiveness of the proposed algorithm. On real-life instances, different batching policies of the samples are evaluated. The more reliable batching policy, in terms of feasibility, is to use a time span of 45 minutes for grouping samples, while a time span of 60 minutes can be better if batches are created with more samples but longer lifetimes, as explained in Section \ref{sec:LUM:inst}. The results also  show that the ALNS algorithm is able to find solutions in which all the samples are delivered on time, while in the real case of Bologna, about 40\% of samples is delivered late.


\section*{Acknowledgements}
The authors are grateful to Aldo Bonadies and Rita Mancini, from the 
Local  Health Care Authority of Bologna, for their collaboration to the problem definition and to provide the data used in the experimentation.


\begin{thebibliography}{10}

\bibitem{ods2019}
M.~Benini, P.~Detti, and G.~Zabalo Manrique~de Lara.
\newblock A milp model for biological sample transportation in healthcare.
\newblock {\em In: M. Paolucci et al. (eds), Advances in Optimization and
  Decision Science for Society, AIRO Springer Series}, 3:81--94, 2019.

\bibitem{Bonadies}
A.~Bonadies, R.~Mancini, M.~Maci, C.~Gibertoni, and A.~Petrini.
\newblock Laboratorio unico metropolitano: innovazione e alta tecnologia per un
  nuovo paradigma di medicina di laboratorio.
\newblock {\em MECOSAN}, XXIX(115):79--94, 2020.

\bibitem{cortes2010pickup}
C.~E. Cort{\'e}s, M.~Matamala, and C.~Contardo.
\newblock The pickup and delivery problem with transfers: Formulation and a
  branch-and-cut solution method.
\newblock {\em European Journal of Operational Research}, 200(3):711--724,
  2010.

\bibitem{DPZomega}
P.~Detti, F.~Papalini, and G.~Zabalo Manrique~de Lara.
\newblock A multi-depot dial-a-ride problem with heterogeneous vehicles and
  compatibility constraints in healthcare.
\newblock {\em Omega}, 70:1--14, 2017.

\bibitem{ctw20}
P.~Detti, G.~Zabalo Manrique~de Lara, and M.~Benini.
\newblock A metaheuristic apporach for biological sample transportation in
  healthcare.
\newblock {\em In: G. Gentile et al. (eds), Graphs and Combinatorial
  Optimization: from Theory to Applications, AIRO Springer Series}, 5:81--94,
  2021.

\bibitem{Doerner2008}
K.~Doerner, M.~Gronalt, R.~Hartl, G.~Kiechle, and M.~Reimann.
\newblock Exact and heuristic algorithms for the vehicle routing problem with
  multiple interdependent time windows.
\newblock {\em Computers \& Operations Research}, 9(35):3034 -- 3048, 2008.

\bibitem{grasas2014improvement}
A.~Grasas, H.~Ramalhinho, L.~S. Pessoa, M.~G. Resende, I.~Caball{\'e}, and
  N.~Barba.
\newblock On the improvement of blood sample collection at clinical
  laboratories.
\newblock {\em BMC health services research}, 14(1):12, 2014.

\bibitem{karakoc2017priority}
M.~Karakoc and M.~Gunay.
\newblock Priority based vehicle routing for agile blood transportation between
  donor/client sites.
\newblock In {\em Computer Science and Engineering (UBMK), 2017 International
  Conference on}, pages 795--799. IEEE, 2017.

\bibitem{liu}
R.~Liu, X.~Xie, and T.~Garaix.
\newblock Hybridization of tabu search with feasible and infeasible local
  searches for periodic home health care logistics.
\newblock {\em Omega}, 47:17--32, 2014.

\bibitem{masson2013adaptive}
R.~Masson, F.~Lehu{\'e}d{\'e}, and O.~P{\'e}ton.
\newblock An adaptive large neighborhood search for the pickup and delivery
  problem with transfers.
\newblock {\em Transportation Science}, 47(3):344--355, 2013.

\bibitem{mitrovic2006pickup}
S.~Mitrovi{\'c}-Mini{\'c} and G.~Laporte.
\newblock The pickup and delivery problem with time windows and transshipment.
\newblock {\em INFOR: Information Systems and Operational Research},
  44(3):217--227, 2006.

\bibitem{pisinger2007}
D.~Pisinger and S.~Ropke.
\newblock A general heuristic for vehicle routing problems.
\newblock {\em Computers \& operations research}, 34(8):2403--2435, 2007.

\bibitem{potvinrosseau}
J.-Y. Potvin and J.-M. Rousseau.
\newblock A parallel route building algorithm for the vehicle routing and
  scheduling problem with time windows.
\newblock {\em European Journal of Operational Research}, 66(3):331--340, 1993.

\bibitem{rais2014}
A.~Rais, F.~Alvelos, and M.~Carvalho.
\newblock New mixed integer-programming model for the
  pickup-and-deliveryproblem with transshipment.
\newblock {\em Eur. J. of Op. Res.}, 3(235):530--539, 2014.

\bibitem{ropke2006}
S.~Ropke and D.~Pisinger.
\newblock An adaptive large neighborhood search heuristic for the pickup and
  delivery problem with time windows.
\newblock {\em Transportation science}, 40(4):455--472, 2006.

\bibitem{csahinyazan2015selective}
F.~G. {\c{S}}ahinyazan, B.~Y. Kara, and M.~R. Taner.
\newblock Selective vehicle routing for a mobile blood donation system.
\newblock {\em European Journal of Operational Research}, 245(1):22--34, 2015.

\bibitem{sapountzis1990allocating}
C.~Sapountzis.
\newblock Allocating blood to hospitals as a multiobjective transportation
  problem.
\newblock In {\em Medical Informatics Europe’90}, pages 733--739. Springer,
  1990.

\bibitem{Shaw}
P.~Shaw.
\newblock Using constraint programming and local search methods to solve
  vehicle routing problems.
\newblock {\em In: Maher M, Puget JF, eds. Proc. 4th Internat. Conf. Principles
  and Practice of Constraint Programming}, 3:417--431, 1998.

\bibitem{Zhang}
Z.~Zhang, M.~Liu, and A.~Lim.
\newblock A memetic algorithm for the patient transportation problem.
\newblock {\em Omega}, 54:60--71, 2015.

\end{thebibliography}
\end{document}